\documentclass[a4paper]{aa}
\usepackage{natbib}
\usepackage{graphicx}
\usepackage{epsfig}
\usepackage{amsmath,amssymb}

\newcommand{\hdos}{$H_{2}$}
\def\M#1{{\mathbf{#1}}}	
\def\T#1{{{#1}^{\mbox{\scriptsize T}}}} 
\def\mdot{\!\cdot\!}		

\begin{document}
\def\sizex{16.0 cm}
\def\bigx{10.0 cm}
\def\smallerxsize{7.0 cm}
\def\smallxsize{10.0 cm}
\def\smallysize{12.0 cm}

\title{Relative abundance pattern along the profile of high redshift
Damped Lyman-$\alpha$ systems\thanks{Based on observations carried out 
at the European Southern  Observatory (ESO), under progs. ID 65.O-0063,
66.A-0624, 67.A-0078, 68.A-0106 and 68.A-0600, with the UVES 
echelle spectrograph installed at the Very Large Telescope (VLT)
unit KUEYEN on Mount Paranal, Chile.}}
\titlerunning{Relative abundance pattern along the profile of DLAs}

\author{E. Rodr\'iguez \inst{1} \and P. Petitjean \inst{1,2} \and B. Aracil
  \inst{1,3} \and C. Ledoux \inst{4} \and R. Srianand \inst{5}}

\offprints {E. Rodr\'iguez, rodrig@iap.fr}

\date{}
\institute{Institut d'Astrophysique de Paris, UMR 7095, 98 bis Boulevard Arago,
  75014 Paris, France \and Observatoire de Paris, LERMA, UMR 8112, 61 Avenue de
  l'Observatoire, 75014 Paris, France \and Department of Astronomy,
  University of Massachusetts, 710 North Pleasant Street, Amherst, MA
  01003-9305 \and European Southern
  Observatory, Alonso de Cordova 3107, Casilla 19001, Vitacura,
  Santiago, Chile \and IUCAA, Post Bag 4,  Ganesh Khind, Pune 411 007,
  India} 

\abstract{We investigated abundance ratios along the profiles of
six high-redshift Damped Lyman-$\alpha$ systems, three of them
associated with H$_2$ absorption, and derived optical depths in each velocity pixel.
The variations of the pixel abundance ratios were found to be remarkably small
and usually smaller than a factor of two within a profile. This result
holds even
when considering independent sub-clumps in the same system. The depletion
factor is significantly enhanced only in those components where H$_2$ is detected.
There is a strong correlation between [Fe/S] and [Si/S] abundances ratios,
showing that the abundance ratio patterns are definitely 
related to the presence of dust. The depletion pattern is usually
close to the one seen in the warm halo gas of our Galaxy.

  }

\maketitle

\section{Introduction}\label{sec:introduction}

Damped Lyman-$\alpha$ systems (hereafter DLAs) observed 
in QSO spectra are characterized by strong H~{\sc i}$\lambda$1215 
absorption lines with broad damping wings.
Although the definition has been restricted for historical reasons 
to absorptions with 
log~$N$(H~{\sc i})~$>$~20.3 (Wolfe et al. 1986), 
damping wings are easily detected in present-day, high quality data for much
lower column densities (down to log~$N$(H~{\sc i})~$\sim$~18.5). 
A more appropriate definition should be related to the physical
state of the gas. If we impose the condition that the
gas must be neutral, then the definition should be limited to systems with
log $N$(H~{\sc i})~$>$~19.5 (e.g. Viegas 1995).

Since their discovery twenty years ago \citep{Wolfe86},
DLAs clearly have something to do with galaxy formation.
What kind of galaxy DLAs are associated to is, however, still a matter of debate.
Some authors identify these systems with large rotating discs \citep{Prochaska97,Hou01},
while others think that DLAs arise mostly either in dwarf galaxies
\citep{Centurion} or galactic building blobs \citep{Haehnelt98,Ledoux98}.
The answer is probably not unique.
In any case, DLAs represent the major reservoir of neutral hydrogen at any redshift
\citep{Storrie2000}, and they probe the chemical enrichment and evolution
of the neutral Universe  (see Pettini et al. 1994; Lu et al. 1996;
Prochaska et al. 1999; Ledoux et al. 2002a and references therein).
Since abundances can be measured very accurately
in DLAs, we can both discuss the connection
between observed abundance ratios and dust content and 
to trace the nucleosynthesis history of the dense gas in the universe.

In this context, it is helpful to compare these results with measurements 
in the ISM of our Galaxy. 
Refractory elements that condense easily into dust grains - namely, Cr,
Fe, Ni- are strongly depleted (up to a factor hundred) in the ISM, while non-refractory 
elements remain in its gaseous phase - S, Zn, and partially Si-. The amount of
depletion depends on the physical condition of the gas. Thus, different depletion 
patterns are observed depending on whether the gas is cold or warm and/or whether
the gas is located in the disc or the halo of the Galaxy
\citep{Savage96}. The LMC and SMC also exhibit different gas-phase abundance 
ratios \citep{Welty99}.

However, a particular nucleosynthesis history can give rise to peculiar
metallicity patterns and mimic the presence of dust. 
\citet{Tinsley} suggested that type Ia supernova are the
major producers of Fe. An enhancement in [$\alpha$/Fe] ratios
($\alpha$-elements are mostly O, S, Si) could reflect an IMF skewed to high masses and
therefore a predominant role of type II supernova. 
For very low metallicity stars ([Fe/H]~$<$~$-$3) in
the Galaxy,  large variations in several abundance ratios have been
reported \citep{william}, which suggests that peculiar
nucleosynthesis processes and inhomogeneous chemical enrichment are probably
taking place. 

As mentioned above, DLAs trace the chemical evolution of galaxies at early
epochs in the universe. Many  detailed studies have been performed so far,
revealing that their metallicities range between 1/300
$Z_{\odot}$ and solar values. The abundance pattern is fairly uniform and 
compatible with low dust content 
(see Pettini et al. 1994, Lu et al. 1996, Prochaska et al. 1999,
Ledoux et al. 2002a). This uniformity in the relative abundance patterns observed
from one DLA to the other has been emphasized by Prochaska \& Wolfe (2002) 
and suggests that protogalaxies have common enrichment histories.

Few studies have adressed the question of the homogeneity inside each
particular system. 
Prochaska \& Wolfe (1996) first studied chemical abundance 
variations in a single DLA, and showed that the chemical abundances were 
uniform to within statistical uncertainties. Lopez et al. (2002)
confirmed this finding from analysis of another DLA using Voigt profile
decomposition.
Petitjean et al. (2002) and Ledoux et al. (2002b) showed that the depletion 
patterns in subcomponents were very similar along DLA profiles 
except in the components where molecular hydrogen is detected 
and where depletion is larger.
More recently, \citet{Prochaska03}, performed a study of 13 systems concluding 
that the majority of DLAs have very uniform relative abundances. This contrasts
in particular with the dispersion in nucleosynthetic enrichment of the Milky Way as 
traced by stellar abundances.
Here, we use the best data from our survey of DLAs (Ledoux et al. 2003) to 
investigate  this issue further using an inversion method to derive the 
velocity profiles in different abundance ratios. In particular, we
investigate the consequence of the presence of molecular hydrogen in 
some of the DLAs.
The paper is structured  as follows: we describe the data in Sect.
2; in Sect. \ref{sec:method} we briefly introduce the method used
for the analysis; and results are presented and discussed in Sects.
\ref{sec:results} and \ref{sec:discussion}.

%
\section{Observations and sample}\label{sec:obs}
\begin{table*}
  \centering
  \begin{tabular}{lcccccc}\hline
    Quasar & $z_{\rm em}$ & $z_{\rm abs}$ & log$N$(H~{\sc i})$^{\rm a}$ & log$N$(H$_{2}$) &
    [Fe/H] &   Metallicity      \\\hline
    $0013-004$ & 2.09 & 1.973 & 20.83 & 17.72 & $-1.46\pm0.01$ & $-0.68\pm0.03$\\
    $0405-443$ & 3.00 & 2.595 & 20.90 & 18.16 & $-1.30\pm0.10$ & $-0.93\pm0.02$\\
    $0405-443$ & 3.00 & 2.549 & 21.00 & $<13.9$ & $-1.49\pm0.01$ & $-0.96\pm0.04$\\
    $0528-250$ & 2.80 & 2.811 & 21.35 & 18.22 & $-1.44\pm0.01$ & $-1.07\pm0.01$\\
    $1037-270$ & 2.31 & 2.139 & 19.70 & $<13.7$ & $-0.51\pm0.03$ & $-0.19\pm0.02$ \\
    $1157+014$ & 1.99 & 1.944 & 21.80 & $<14.5$ & $-1.84\pm0.03$ & $-1.41\pm0.02$\\
  \end{tabular}
  \caption{Description of the sample. The neutral and molecular hydrogen column 
densities were taken from Ledoux et al. (2003). Both [Fe/H] and metallicity values
  were from this work and correspond to the values integrated over the 
  systems. Metallicity was derived in all cases from [S/H], 
  except for $Q~1157+014$ for which [S/H] determination was uncertain,
  so we  used [Zn/H] instead.}
  \label{tab:obs}
\end{table*}

As emphasized by Prochaska (2003), a high S/N ratio is needed to 
investigate variations along the profile.
In addition, we want to investigate the difference between systems
where molecular hydrogen is and where is not detected
because it has been shown (e.g. Petitjean et al. 2002) that depletion 
is larger in components where H$_2$ is detected. 
We therefore restricted the sample to the six highest S/N ratio spectra
from the list of 33 quasars (24 DLAs and 9 sub-DLAs) observed during the VLT survey
for molecular hydrogen in DLAs (Ledoux et al. 2003).
The QSOs were observed with the
Ultraviolet and Visible Echelle Spectrograph \citep{dodorico}, mounted  on
the 8.2 m Kueyen telescope operated at Cerro Paranal, Chile, during the
observation periods P65 to P68. The actual spectral resolution lies
in the range 42500$<$R$<$53000.
For all the spectra, the S/N ratio is larger than 50 per pixel 
(see Ledoux et al. 2003 for details). For Q~0528$-$250,
we also retrieved complementary data from the ESO archive, thereby obtaining a 
very high-quality spectrum with SNR~$\sim100$. Molecular hydrogen was detected in three of these systems.  This is the case for Q~0013$-$004
\citep{ppjq0013}, Q~0528$-$250 \citep{anandq0528}, and Q~0405$-$443 -system
at $z_{abs}=2.595$- (Ledoux et al. 2003). 
The six systems span a wide range of H~{\sc i} column densities from 
log $N$(H~{\sc i}) = 19.7 up to 21.80.

In all systems we chose to analyze absorption features that are well-defined 
and do not suffer from major blending.


\subsection{Q~0013$-$004}\label{subsec:q0013_desc}

\begin{figure}[tbh]
  \centering
  \includegraphics[height=11cm,width=8.5cm]{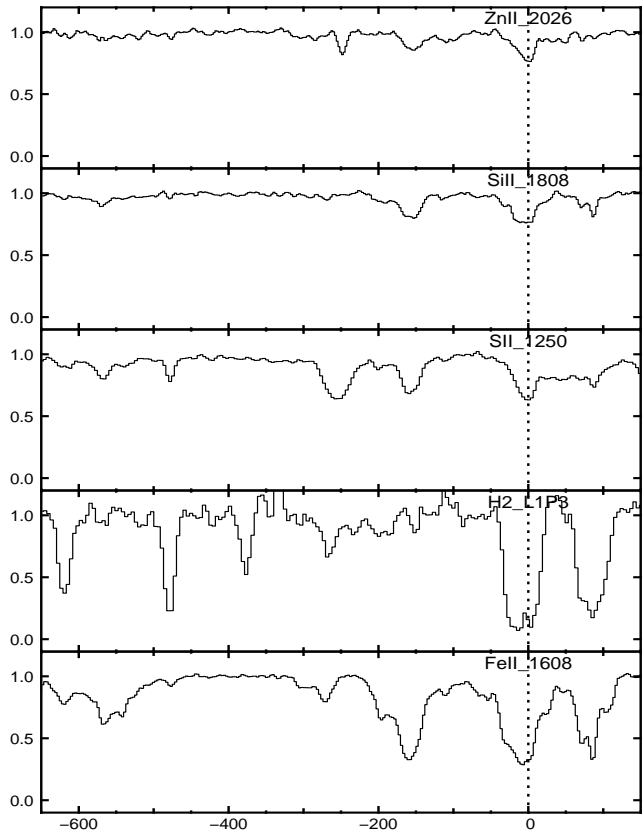}
  \caption{Absorption spectra plotted on a velocity scale for
a few transitions -including H2L1P3- observed in the
  $z_{\rm abs}$~=~1.973 DLA system toward Q~0013$-$004. Molecular 
transitions have been detected in four components located
at $v\sim-615$~km~s$^{-1}$, $v\sim-480$~km~s$^{-1}$,
  $v\sim0$~km~s$^{-1}$,  and $\sim85$~km~s$^{-1}$
  \citep{ppjq0013}.}
  \label{fig:lines_q0013}
\end{figure}

The presence of H$_{2}$ at $z_{\rm abs}=1.973$  was first reported by \citet{Ge}.
Spread over more than 1000~km~s$^{-1}$ (see Fig. \ref{fig:lines_q0013}), 
this system is the DLA absorber with
the highest molecular fraction known so far ($-2.7<f(H_{2})<-0.64$,
Petitjean et al. 2002) for log~$N$(H~{\sc i})~=~20.83.
Molecular hydrogen has been detected in four different components
 spread all over the system, at relative velocities,
 $\sim-615$, $\sim-480$, $\sim0$ and
 $\sim+80$~km~s$^{-1}$. \citet{ppjq0013} also report four additional
 strong metal  components that probably make a non-negligible contribution to the
 total H~{\sc i} column density. All components have similar abundance ratios and
depletion factors independent of the presence or absence of H$_2$. 
Only for the special molecular component at $z_{\rm abs}=1.97296$ does the  dust
depletion turn to be important. 
The  depletion of Fe is comparable to that observed in the
cold interstellar medium of the Galactic disc. We summarize  the overall abundances of the principal elements present in the system in Table \ref{tab:ALLtotalN}. We also confirm the strong depletion pattern in the $\sim480$~km~s$^{-1}$ component.


\subsection{Q~0405$-$443}

Lopez et al. (2001) first discovered three DLA systems along the line of sight
to this quasar at redshifts $z_{\rm abs}$~=~2.550, 2.595 and 2.621. 
Later, \citet{Ledouxsurvey} confirmed the damped nature of these three 
absorptions. We adopted their H~{\sc i} column density values. 
From the three systems observed
in this line of sight, we included the systems at
$z_{\rm abs}$~=~2.549, log~$N$(H~{\sc i})~=~21.0 and $z_{\rm abs}$~=~2.595,
log~$N$(H~{\sc i})~=~20.9 in our sample. In the third system, metallicity is much
lower, [Fe/H]~=~$-$2.15 , which renders the analysis much more uncertain. 
\citet{Ledouxsurvey} detect the presence of H$_2$ at $z_{\rm abs}$~=~2.595 from
J~=~0, 1, 2, 3 rotational transitions. 
They measure log~$N$(H$_2$)~=~18.16, one of the largest H$_2$
column densities ever seen in  DLAs, although the molecular fraction in
the corresponding cloud is not all that large (log~$f$(H$_2$)~=~$-$2.44) due to the 
high column density of neutral hydrogen.


\begin{figure}[tbh]
  \centering
  \includegraphics[height=11cm,width=8.5cm]{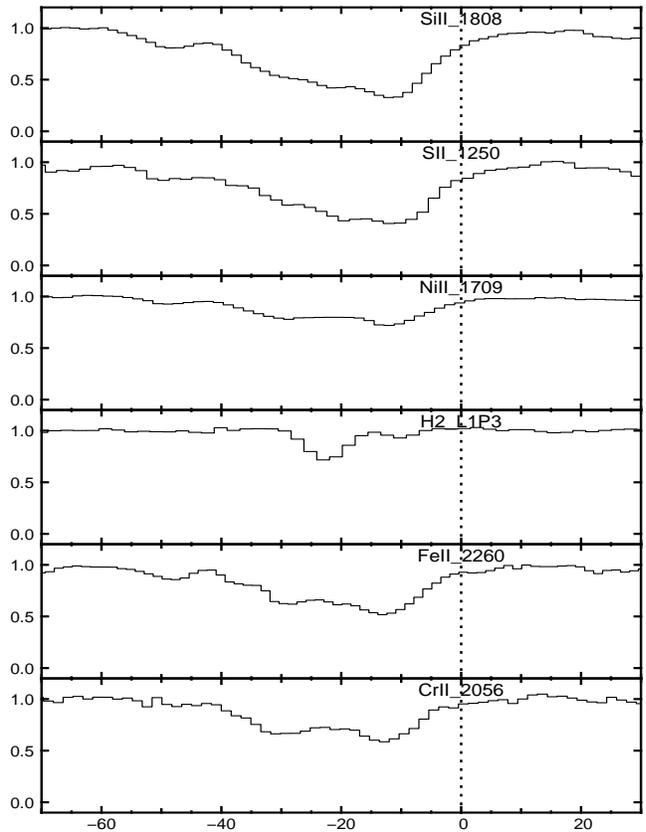}
  \caption{Absorption spectra plotted on a velocity scale for
a few transitions -including H2L1P3 transition-
observed in the
  $z_{\rm abs}$~=~2.595 DLA system toward Q~0405$-$443. Numerous molecular 
transitions have been detected in two components located
at $v\sim-24$~km~s$^{-1}$ and $\sim-11$~km~s$^{-1}$
  \citep{Ledouxsurvey}.}
  \label{fig:lines_q0405}
\end{figure}

\begin{figure}[tbh]
  \centering
  \includegraphics[height=11cm,width=8.5cm]{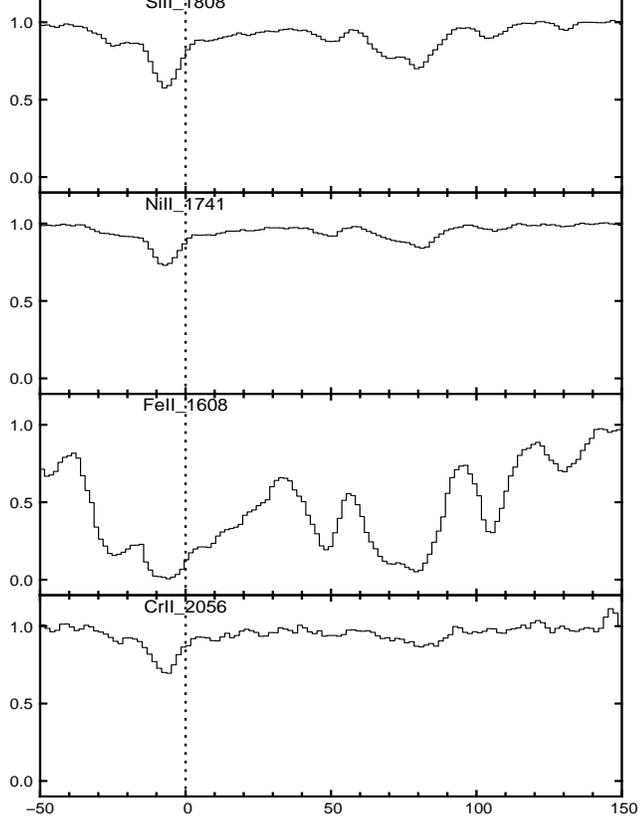}
  \caption{Absorption spectra plotted on a velocity scale for
a few transitions 
observed in the
  $z_{\rm abs}$~=~2.55 DLA system toward Q~0405$-$443.
  \citep{Ledouxsurvey}}
  \label{fig:lines_q0405b}
\end{figure}


\subsection{PKS~0528$-$250}

\begin{figure}[tbh]
  \centering
  \includegraphics[height=11cm,width=8.5cm]{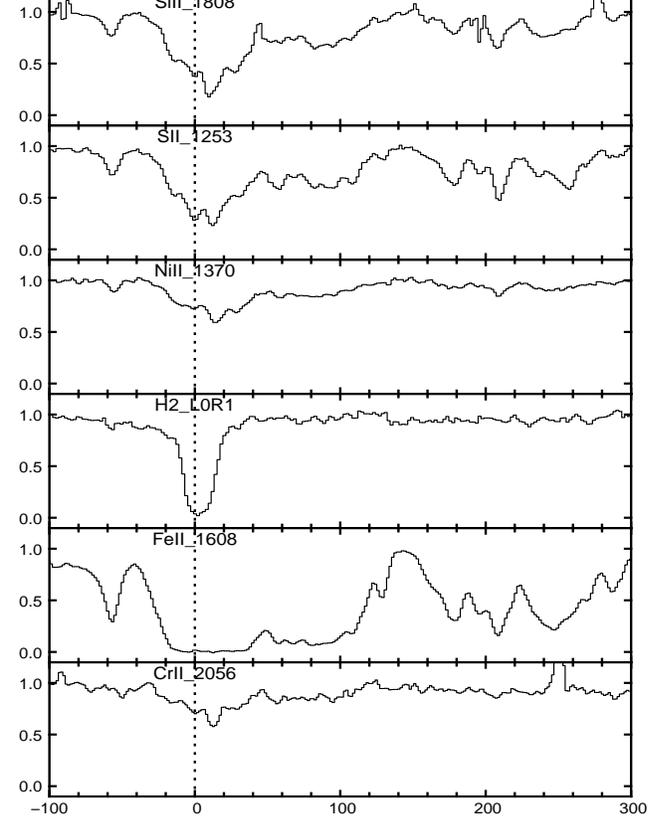}
  \caption{Absorption spectra plotted on a velocity scale for
a few low-ionization species -including H2L0R1 transition-
  observed at $z_{\rm abs}~=~2.811$ along the line of sight to 
  PKS~0528$-$250. The system extends over more than 400~km~s$^{-1}$. }
  \label{fig:lines_q0528}
\end{figure}

We collected complementary data from the ESO archive available on this QSO
and added them together with our own data. The result is a
very high SNR spectrum that extends over the wavelength range 3000$-$10000~\AA~, 
apart from a few gaps.
The absorption redshift is slightly higher than that of the emitting
source, suggesting that the absorbing system could be associated with the
quasar. A useful consequence of this is that a large number of 
metallic transitions are redshifted outside the Lyman-$\alpha$ forest
and therefore can be used for our analysis.
This system has been known for many years to be the only system at high redshift 
where molecules were detected \citep{Varsha85,Cowie96,anandq0528}.
New data have been obtained with VLT by Ledoux et al. (2003).
These authors derive log~$N$(H$_2$)~=~17.93 and 18.0 in two components
at $z_{\rm abs}$~=~2.81100 and 2.81112, respectively. Given 
the large neutral  hydrogen column density (log~$N$(H~{\sc i})~=~21.35), the molecular
fraction is only $f$(H$_{2}$)~=~9$\times$10$^{-4}$. The excitation temperature
for the J~=~1 rotational level is between 150 and 200~K, and the
density is probably quite large (Srianand \& Petitjean 1998).
As mentioned by \citet{Lu96}, the metal absorption lines are
unusually wide and complex, spreading over about 400~km~s$^{-1}$. They appear to 
be structured in two main sub-clumps above and below +140~km~s$^{-1}$ (see 
Fig.~\ref{fig:lines_q0528}).


\subsection{Q~1037$-$270}


\begin{figure}[tbh]
  \centering
  \includegraphics[height=11cm,width=8.5cm]{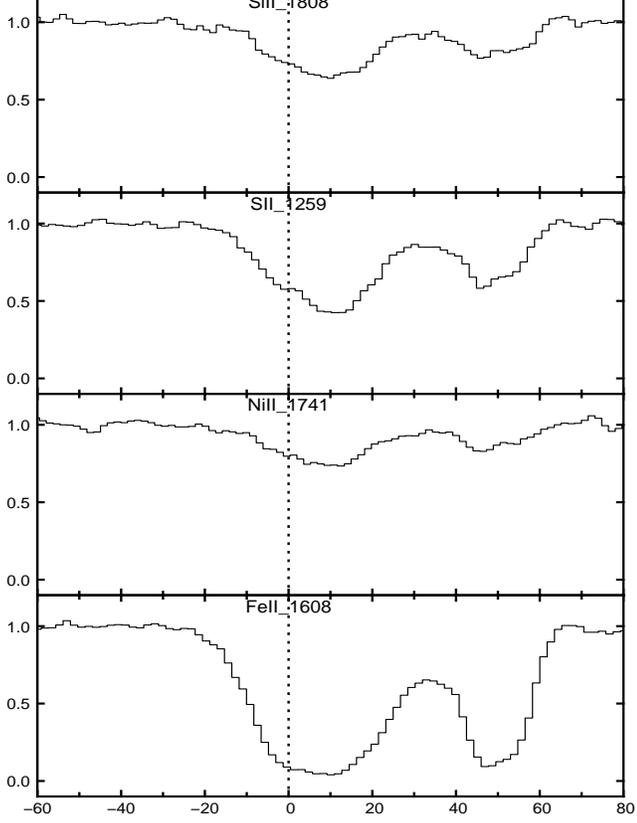}
  \caption{Absorption spectra plotted on a velocity scale
for selected transition lines in the  DLA
  system at  $z_{\rm abs}=2.139$ towards Q~1037$-$270. The
  system shows  a ``smooth'' profile for all the elements spread over
  nearly 100~km~s$^{-1}$}
  \label{fig:lines_q1037}
\end{figure}


The continuum of the QSO is difficult to constrain due to
the presence of a complex system of BAL troughs (see Srianand
\& Petitjean 2001). 
A first limit on the neutral hydrogen
column density was derived by  Lespine \& Petitjean (1997) from the absence of
damped wings in a low-resolution spectrum. 
We adopted log $N$(H~{\sc i})~=~19.7 from Srianand \& Petitjean (2001). 
This is the lowest column density in our sample. However,
this system  has the highest metallicity known for DLAs at such a redshift
($z_{\rm abs}=2.139$), [Zn/H]~=~$-$0.26.
As a consequence of this high metallicity, many low-ionization lines were detected,
including C~{\sc i}, though no molecules were detected. Some of the corresponding
species are shown in Fig.~\ref{fig:lines_q1037}.
The depletion pattern derived by  Srianand \& Petitjean (2001) is compatible
with very low dust content, if any at all.
The gas seems to be warm and halo-like. 

\subsection{Q~1157+014}

This system has the highest H~{\sc i} column density in our sample
(log~$N$(H~{\sc i})~=~21.8~cm$^{-2}$) and is close to the
emission redshift of the quasar.
Absorption in 21~cm has been detected by \citet{Briggs84}
and the spin temperature is constrained by Kanekar \& Chengalur (2003) 
to be 865$\pm$190~K. Neither H$_2$ nor C~{\sc i} were detected
and the metallicity is the smallest in the sample 
([Zn/H]~=~$-$1.41, see Table~1).


\begin{figure}
  \centering
  \includegraphics[height=11cm,width=8.5cm]{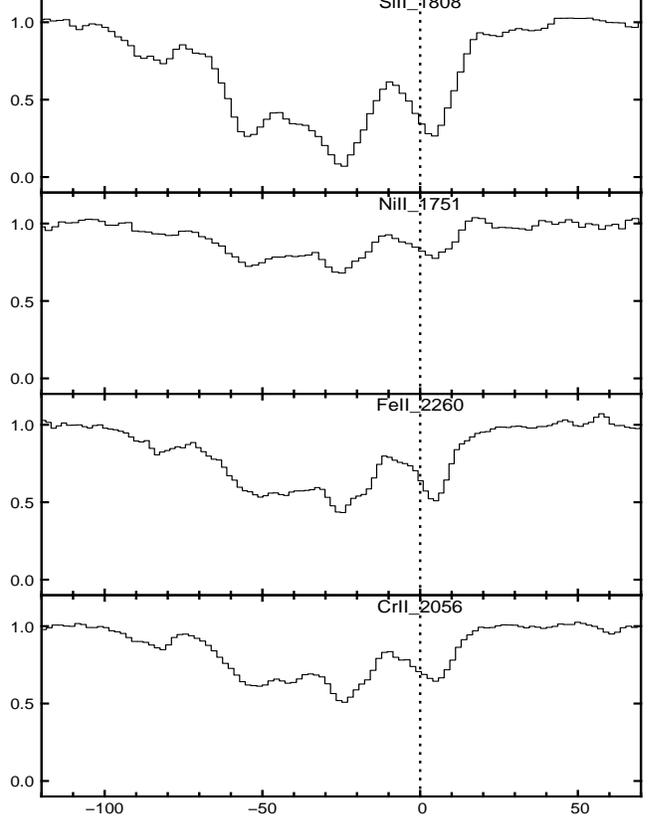}
  \caption{Absorption spectra plotted on a velocity scale
for a few transitions observed in the DLA
  system at $z_{\rm abs}=1.944$ towards Q~1157+014.}
 \label{fig:lines_q1157}
\end{figure}

\begin{table}
\caption{Atomic data}\label{tabosc}
\begin{tabular}{llll}
\hline
\hline
Transition                & $\lambda _{\rm vac}$ & $f$    & Ref.\\
                          & (\AA )               &        &     \\
\hline
Si\,{\sc ii}\,$\lambda$1020   & 1020.6989 & 0.01680  & a \\
Si\,{\sc ii}\,$\lambda$1190   & 1190.4158 & 0.25000  & a \\
Si\,{\sc ii}\,$\lambda$1304   & 1304.3702 & 0.08630  & a \\
Si\,{\sc ii}\,$\lambda$1526   & 1526.7070 & 0.12700  & a \\
Si\,{\sc ii}\,$\lambda$1808   & 1808.0129 & 0.00208  & a \\
S\,{\sc ii}\,$\lambda$1250    & 1250.5840 & 0.00545  & a \\
S\,{\sc ii}\,$\lambda$1253    & 1253.8110 & 0.01090  & a \\
S\,{\sc ii}\,$\lambda$1259    & 1259.5190 & 0.01620  & a \\
Cr\,{\sc ii}\,$\lambda$2056   & 2056.2690 & 0.1050    & a \\
Cr\,{\sc ii}\,$\lambda$2062   & 2062.2361 & 0.0780   & a \\
Cr\,{\sc ii}\,$\lambda$2066   & 2066.1640 & 0.0515   & a \\
Fe\,{\sc ii}\,$\lambda$1143   & 1143.2260 & 0.01770  & a \\
Fe\,{\sc ii}\,$\lambda$1144   & 1144.9379 & 0.10600 & a \\
Fe\,{\sc ii}\,$\lambda$1608   & 1608.4508 & 0.05800  & a \\
Fe\,{\sc ii}\,$\lambda$1611   & 1611.2003 & 0.00136 & a \\
Fe\,{\sc ii}\,$\lambda$2249   & 2249.8768 & 0.00182  & a \\
Fe\,{\sc ii}\,$\lambda$2260   & 2260.7805 & 0.00244  & a \\
Fe\,{\sc ii}\,$\lambda$2344   & 2344.2130 & 0.11400 & a \\
Fe\,{\sc ii}\,$\lambda$2374   & 2374.4603 & 0.0313   & a \\
Ni\,{\sc ii}\,$\lambda$1370   & 1370.1320 & 0.0765   & b \\
Ni\,{\sc ii}\,$\lambda$1454   & 1454.8420 & 0.0300   & b \\
Ni\,{\sc ii}\,$\lambda$1709   & 1709.6042 & 0.0324   & b \\
Ni\,{\sc ii}\,$\lambda$1741   & 1741.5531 & 0.0427   & b \\
Ni\,{\sc ii}\,$\lambda$1751   & 1751.9157 & 0.0277   & b \\
Zn\,{\sc ii}\,$\lambda$2026   & 2026.1371 & 0.489    & a \\
Zn\,{\sc ii}\,$\lambda$2062   & 2062.6604 & 0.256    & a \\
\hline
\hline
\end{tabular}
\flushleft {\sc References:} Vacuum wavelengths from Morton (2003).\\ Oscillator strengths: (a)~Morton (2003); (b)~Fedchak et al. (2000).
\end{table}


\section{Analysis method}\label{sec:method}

We want to estimate the column density per unit velocity
along the absorption profile of species X  using
several transitions.   
Following Savage \& Sembach (1991), we can write that the 
apparent optical depth per unit velocity of species X in the QSO spectrum 
at wavelength $\lambda$ for any transition of oscillator strength
$f$  and rest wavelength $\lambda_0$ is
$\tau_{\rm a}(\lambda)$~=~Ln($1/F(\lambda)$), and the column density
per unit velocity is 
log $N_{\rm a}(\lambda)$~=~log~$\tau_{\rm a}(\lambda)$$-$log~$f\lambda_0$$-$14.976.
Given a set  of $m$ transitions of the same species X, we can use the
duplication  of  the information  over  the  different transition lines  to
derive  the  optical depth  profile  of  the species watching out  
in addition, for
possible  blending.  Let  $\lambda_{\rm 0}^{\rm k}$  and  
$f^{\rm k}$ be, respectively,  the laboratory wavelength and the 
oscillator strength of transition $k$.
First, the regions corresponding to these transitions are rebinned to
the smallest pixel size so that they span the same redshift range over
an identical  number $n$ of  pixels.  Let $\lambda_{\rm i}^{\rm k}$,  
$F_{\rm i}^{\rm k}$ be
the wavelength  and the normalized flux  at pixel i  of the region
corresponding     to transition    $k$.      The     value
$\lambda_{\rm i}^{\rm k}/\lambda_{0}^{\rm k}-1$ is  independent of $k$,  
thanks to the rebin, and is the redshift  at pixel $i$.  The observed optical
depth, $-\ln\,F_{\rm i}^{\rm k}$, is considered as  the sum of the 
optical depth, $\tau_{\rm i}^{\rm Xk}$, of transition  $k$  and  the 
optical depth, $\tau_{\rm i}^{\rm Bk}$, of  a possible
intervening absorption blended with the absorption of interest
(of course in practice we will choose profiles that are
apparently not blended with interlopers). 
As the different transitions are from the same species, the
quantity log~$N_{\rm a}^{\rm k}$ (or simply 
$\tau_{\rm i}^{\rm Xk}$/$f^{\rm k}\lambda^{\rm k}_0$)
is the same for all transitions $k$.

The  fitted   optical  depth  before   any  instrumental
convolution  and without  adjusting for any  overlapping  between 
regions, is then:
\begin{equation}
\tau_{\rm i}^{\rm k}=\tau_{\rm i}^{\rm Xk}\,+\,\tau_{\rm i}^{\rm Bk}
\;\;\;\forall i\in\{1,\ldots,n\}\;,\;\forall k\in\{1,\ldots,m\}.\label{eq:tau_fit}
\end{equation}
To simplify  the notation, we use  the data vector  $\M{O}$ of $m*n$
elements, obtained  by stacking  the observed flux  of each  region (i.e.
$F_{\rm i}^{\rm k}=\M{O}_{\rm i+kn-n}$),  the parameter  vector  $\M{P}$ 
of  $n+m*n$ elements,  obtained   by  stacking  
$\tau_{\rm i}^{\rm Xk}$/$f^{\rm k}\lambda^{\rm k}_0$ 
and  $\tau_{\rm i}^{\rm Bk}$ (i.e.
$\tau_{\rm i}^{\rm Xk}$/$f^{\rm k}\lambda^{\rm k}_0=\M{P}_i$  and  
$\tau_{\rm i}^{\rm Bk}=\M{P}_{kn+i}$),  and the  fitted
vector $\M{F}$ corresponding to the fitted flux.  The $\chi^2$
of the fit is then:
\begin{eqnarray}
&\chi^2&=\mbox{tr}\left(\T{[\M{W}\mdot(\M{O}-\M{F})]}\mdot[\M{W}\mdot(\M{O}-\M{F})]\right)\label{eq:chi2}\\
with &\M{F}&=\M{C}\mdot\exp^*(-\M{M}\mdot\M{P}).
\end{eqnarray}
The matrix $\M{M}$ applies the linear relation (\ref{eq:tau_fit}) to the
parameters  and  takes  care  of possible  overlap  between different
regions. The  function $\exp^*$ replaces  each element of a  matrix by
its exponential  value (i.e. $\exp^*(X)_{ij}=e^{X_{ij}}$).  The matrix
$\M{C}$  is  used to  convolve  the calculated flux with  the  
instrumental profile  and,
finally, $\M{W}$  is the inverse  of the variance-covariance  matrix of
the data.

A regularization  constraint is added to select  the solutions $\M{P}$
that are  correlated over a  specific length $l_0$.  The  selection is
done by  a  minimization of the  high frequency coefficients
of the  discrete Fourier transform  of the parameters $\M{P}$,  i.e. by
minimizing $\zeta^2$ (Eq.\ref{eq:regul} below) at the same  time as $\chi^2$ (Eq.\ref{eq:chi2}).
\begin{equation}
  \zeta^2=tr\left(\T{[\M{S}_{l_0}\mdot\M{T}\mdot\M{P}]}\mdot[\M{S}_{l_0}\mdot\M{T}\mdot\M{P}]\right),\label{eq:regul}
\end{equation}
where $\M{T}$  is  the matrix  of  the  discrete  Fourier transform and
$\M{S}_{l_0}$ the high frequency filter that cancels the coefficients
of the Fourier transform corresponding to a length (in the parameter
space) greater than $l_0$.

Actually, the  overall fit with the regularization  constraint is done
by  using a  Lagrange  parameter $f_0$  and  minimizing the  following
quantity,
\begin{equation}
Q=\chi^2\,+\,f_0\,\zeta^2
\end{equation}
The   parameter  $f_0$   controls   the  relative   strength  of   the
regularisation  over the $\chi^2$  fit of  the data.  If $f_0$  is too
small, then only  $\chi^2$ is minimized and
the data will  be over-fitted. In  contrast, if  $f_0$ is too  large, the
minimization is done only for  $\zeta^2$ to give a very smooth solution
that  does not  fit the  data. Thus,  $f_0$ is  chosen to  obtain 
a $\chi^2$ value approximately equal to 1. In the worst case,
since  there are  more  parameters ($n*(m+1)$) than data  points ($n*m$),  
the problem  is underestimated.   Nevertheless,  the  regularization   introduces  a
 constraint    that    diminishes   the    real    number   of    free
 parameters. Moreover, in practice, most of the profiles we use here
are free of blending and, in that case, we impose a zero optical depth
for the intervening absorption.

Once the different profiles have been fitted,  we  compute the ratios
between the most important elements in order to analyze possible variations of 
the ratios along the profile.
The method is well-suited to studying the complex absorption profiles of DLAs.
Indeed, it is well-known that the decomposition in discrete components is not unique
due to the fact that the observed absorption profile is a convolution
of absorptions from density fluctuations in a continuous medium. In addition, 
absorptions from different overdensities can be superimposed by peculiar kinematics, 
making any decomposition in components an ad hoc representation of reality.
The pixel-by-pixel method instead does not assume any model and 
emphasizes deviations from the mean value when a fit tends to smooth such 
deviations out. Although the mean depletion measured by both methods, the pixel-by-pixel 
analysis and the sub-component fitting, is approximatively the same, 
deviations from the mean and in particular locations within the gas, can only be 
revealed by a pixel-by-pixel analysis. In particular, the scatter in the data points 
is a good estimate of the (in)homogeneity of the gas.


\section{Results for individual objects}\label{sec:results}


Several of the systems in the sample are spread over more than 200~km~s$^{-1}$.
The absorption profiles are often clearly structured in several well-detached 
clumps with no or very little absorption inbetween. To investigate whether the 
properties differ from one clump to the other, we performed the
analysis on each of the clumps. Indeed, it is reasonable to believe that 
different part of the DLAs could have different characteristics (e.g.  
Haehnelt et al. 1998).
Column densities of several species, integrated over the system,
are given for  each system in Table~\ref{tab:ALLtotalN}.
Abundance ratios relative to solar are defined as
[X/Y]~$\equiv$~log[$N$(X)/$N$(Y)]~-~log[$N$(X)/$N$(Y)]$_{\odot}$, adopting solar
abundances from \citet{Anders93}. These solar abundances and
typical galactic depletion values are summarized in Table~\ref{tab:solar}.
We computed the quantity [X/Y] in each pixel and its average 
($\overline{[X/Y]}$) over each subclump, 
and results are given in Table~\ref{tab:res}. Each subclump is referred to by the
velocity range over which it is spread relative to the main redshift of the 
system, with a reference $z_{abs}$ taken from \citep{Ledouxsurvey}. Then $\overline{[X/Y]}$, averaged over this velocity range, is given
for each ratio. In each row, the number next to  $\overline{[X/Y]}$ is the
error on the mean ratio ($\sigma=\sqrt{\sum_{i=1}^{n}{\sigma_{i}^2}/n}$),  
and the second number (in italics) is the scatter 
of [X/Y] around the mean calculated over the subclump,
$[\sum{(\overline{[X/Y]}-[X/Y]_{i})^{2}}/n]^{1/2}$.


\begin {table}[!hbt]
\caption{Solar abundances and typical galactic depletion values of elements investigated in
    this paper} \label{tab:solar}
\begin{scriptsize}
\begin{tabular}{lcccc}
\hline
\hline
Ion  &  log(X/H)+12$^{\rm a}$&     & [X/S]$^{\rm b,c}$&  \\
               &                                &Cold disc &Warm disc&Halo \\
\hline
 H\,{\sc i}   & $12.00$ & ...& ...      & ...\\
 Si\,{\sc ii} & $7.55\pm 0.02$ & $-1.3$ &$-0.4$ &$-0.3$\\
S\,{\sc ii}  & $7.27\pm 0.05$ & $+0.0$&$+0.0$&$+0.0$\\
 Cr\,{\sc ii} & $5.68\pm 0.03$ & $-2.1$&$-1.2$&$-0.6$\\
 Mn\,{\sc ii} & $5.53\pm 0.04$ & $-1.5$&$-1.0$&$-0.7$\\
 Fe\,{\sc ii} & $7.51\pm 0.01$ & $-2.2$&$-1.4$&$-0.6$\\
 Ni\,{\sc ii}& $6.25\pm 0.02$ & $-2.2$&$-1.4$&$-0.6$\\
Zn\,{\sc ii} & $4.65\pm 0.02$ & $-0.4$&$-0.2$&$-0.1$ \\

\hline
\hline
\end{tabular}
\end{scriptsize}
\flushleft{$^{\rm a}$  Solar \& Meteorite abundances from Anders
(1993)\\
$^{\rm b}$ [X/H]= log[Z(X)]-log[Z$_\odot$(X)]\\
$^{\rm c}$  Galactic values are from Welty et al. (1999)}

\end {table}


\begin{table*}[p]
  \centering
 \caption{Total column densities and integrated abundance ratios relative to solar
  in the six DLAs of our sample.}\label{tab:ALLtotalN}
  \begin{scriptsize}
    \begin{tabular}{cccc}\hline
      Species & $N$(cm$^{-2}$)       & [X/S] & [X/Fe]\\\\\hline\hline\\
      Q~0013$-$004 &$z_{\rm abs}~=~1.973$&&\\\hline
      FeII & $7.55 \pm 0.01 \times 10^{14}$ &  $-0.78\pm0.03$ & $ 0.00\pm0.00$ \\
      SII & $2.64 \pm 0.08 \times 10^{15}$ &   $0.00\pm0.00$ &  $0.78\pm0.03$ \\
      SiII  & $2.72 \pm 0.20 \times 10^{15}$ &   $-0.27\pm0.08$ &  $0.51\pm0.07$ \\ 
      ZnII & $7.30 \pm 1.12 \times 10^{12}$ &  $0.06\pm0.15$ &  $0.84\pm0.15$ \\
      NiII & $7.57 \pm 0.72 \times 10^{13}$ &  $-0.52\pm0.05$ & $0.26\pm0.04$ \\\\
      \hline
    \end{tabular}
    \begin{tabular}{cccc}\hline
      Species & $N$(cm$^{-2}$)       & [X/S] & [X/Fe]\\\\\hline\hline\\
      Q~0405$-$443 & $z_{\rm abs}~=~2.549$              &  & \\\hline
      FeII & $1.05 \pm 0.01 \times 10^{15}$ & $-0.53\pm0.04$ & $0.00\pm0.00$ \\
      SII & $2.03 \pm 0.07\times 10^{15}$ &  $0.00\pm0.00$ & $0.53\pm0.14$ \\
      SiII  & $2.38 \pm 0.32\times 10^{15}$   & $-0.21\pm0.15$ & $0.32\pm0.14$ \\ 
      CrII & $2.46 \pm 0.76\times 10^{13}$ &  $-0.33\pm0.31$ & $0.20\pm0.31$ \\
      NiII & $7.05 \pm 0.96\times 10^{13}$ &  $-0.44\pm0.14$ & $0.09\pm0.14$ \\\\
      \hline
    \end{tabular}
    \begin{tabular}{cccc}
      Q~0405$-$443 & $z_{\rm abs}~=~2.595$& & \\\hline
      FeII & $1.30 \pm 0.13 \times 10^{15}$ &  $-0.37\pm0.10$ & $0.00\pm0.00$ \\
      SII &$1.75 \pm 0.04\times 10^{15}$  & $0.00\pm0.00$ & $0.37\pm0.10$ \\
      SiII  &$3.52 \pm 0.07\times 10^{15}$   & $0.03\pm0.03$ & $0.40\pm0.10$ \\ 
      CrII & $2.23 \pm 0.07\times 10^{13}$ & $-0.30\pm0.04$  & $0.06\pm0.10$ \\
      NiII & $6.92 \pm 0.08\times 10^{13}$ &  $-0.38\pm0.02$ &
      $-0.02\pm0.10$ \\\\
      \hline
    \end{tabular}
     \begin{tabular}{cccc}
      Q~0528$-$250 &$z_{\rm abs}~=~2.811$&&\\\hline
      FeII & $2.60 \pm 0.03 \times 10^{15}$  &  $-0.37\pm0.01$ & $0.00\pm0.00$ \\
      SII & $3.49 \pm 0.02\times 10^{15}$  &   $0.00\pm0.00$ & $0.37\pm0.01$ \\
      SiII  &  $1.01 \pm 0.05\times 10^{16}$   &  $0.18\pm0.05$ & $0.55\pm0.05$ \\ 
      ZnII & $1.27 \pm 0.26\times 10^{13}$   &   $0.18\pm0.21$ & $0.55\pm0.21$ \\
      CrII & $7.18 \pm 1.45\times 10^{13}$   &  $-0.10\pm0.20$ & $0.17\pm0.20$  \\
      NiII & $1.62 \pm 0.32 \times 10^{14}$  &  $-0.31\pm0.20$ & $0.05\pm0.20$ \\
      \hline
    \end{tabular}
    \begin{tabular}{cccc}
      Q~1037$-$270 & $z_{\rm abs}~=~2.139$& & \\\hline
      FeII & $4.98 \pm 0.14 \times 10^{14}$ & $-0.32\pm0.04$ &  $0.00\pm0.00$ \\
      SII & $6.05 \pm 0.15 \times 10^{14}$ & $ 0.00\pm0.00$ & $0.32\pm0.04$ \\
      SiII  & $1.89 \pm 0.15 \times 10^{15}$ & $0.21\pm0.08$ & $0.53\pm0.09$ \\ 
      CrII & $1.02 \pm 0.58 \times 10^{13}$ &  $-0.18\pm0.57$ &$ 0.14\pm0.57$ \\
      NiII & $6.43 \pm 1.04 \times 10^{13}$ &  $0.05\pm0.16$ & $0.36\pm0.16$  \\
      MnII & $2.22 \pm 0.08 \times 10^{12}$ &  $-0.70\pm0.04$ & $-0.37\pm0.05$ \\
      \hline
    \end{tabular}
     \begin{tabular}{cccc}

      Q~1157+014 & $z_{\rm abs}~=~1.944$& [X/Zn]& \\\hline
      FeII & $2.95 \pm 0.08 \times 10^{15}$ &  $-0.43\pm0.04$ &  $0.00\pm0.00$ \\
      SiII & $1.00 \pm 0.01 \times 10^{16}$ &  $0.06\pm0.02$ &  $0.50\pm0.03$ \\
      ZnII & $1.09 \pm 0.02 \times 10^{13}$ &  $0.00\pm0.00$ &  $0.43\pm0.04$ \\
      CrII & $6.07 \pm 0.19 \times 10^{13}$ &  $-0.29\pm0.04$ &  $0.14\pm0.04$ \\
      NiII & $1.62 \pm 0.05 \times 10^{14}$ &  $-0.43\pm0.04$ & $0.00\pm0.04$ \\\\
      \hline
    \end{tabular}
  \end{scriptsize}

\end{table*}

\subsection{Q~0013$-$004}\label{subsec:q0013_res}

\begin{figure*}
  \centering
  \includegraphics[height=10cm,width=12cm]{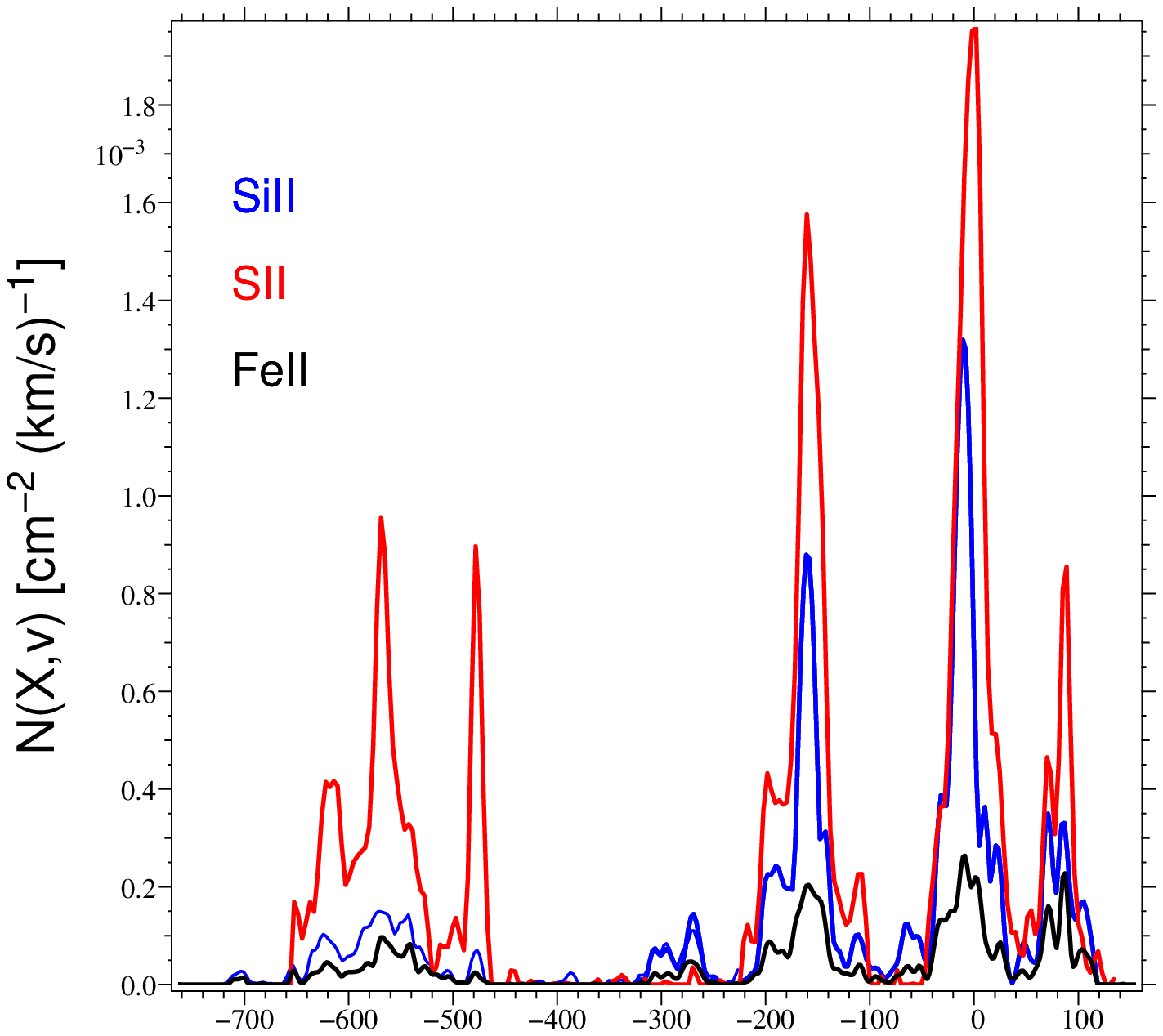}
\caption{This panel shows the optical depth profiles in 
the $z_{\rm abs}=1.973$ DLA toward
    Q~0013$-$004. The apparent column density per velocity bin
    -$N(X,dv)$ (y axis values are scaled to $\times10^{-3}$)- 
is represented on a velocity scale, with $v=0$~km~s$^{-1}$ centered
    at $z_{\rm abs}=1.973$.}\label{fig:q0013a}
 \includegraphics[height=12cm,width=6.2cm,angle=270]{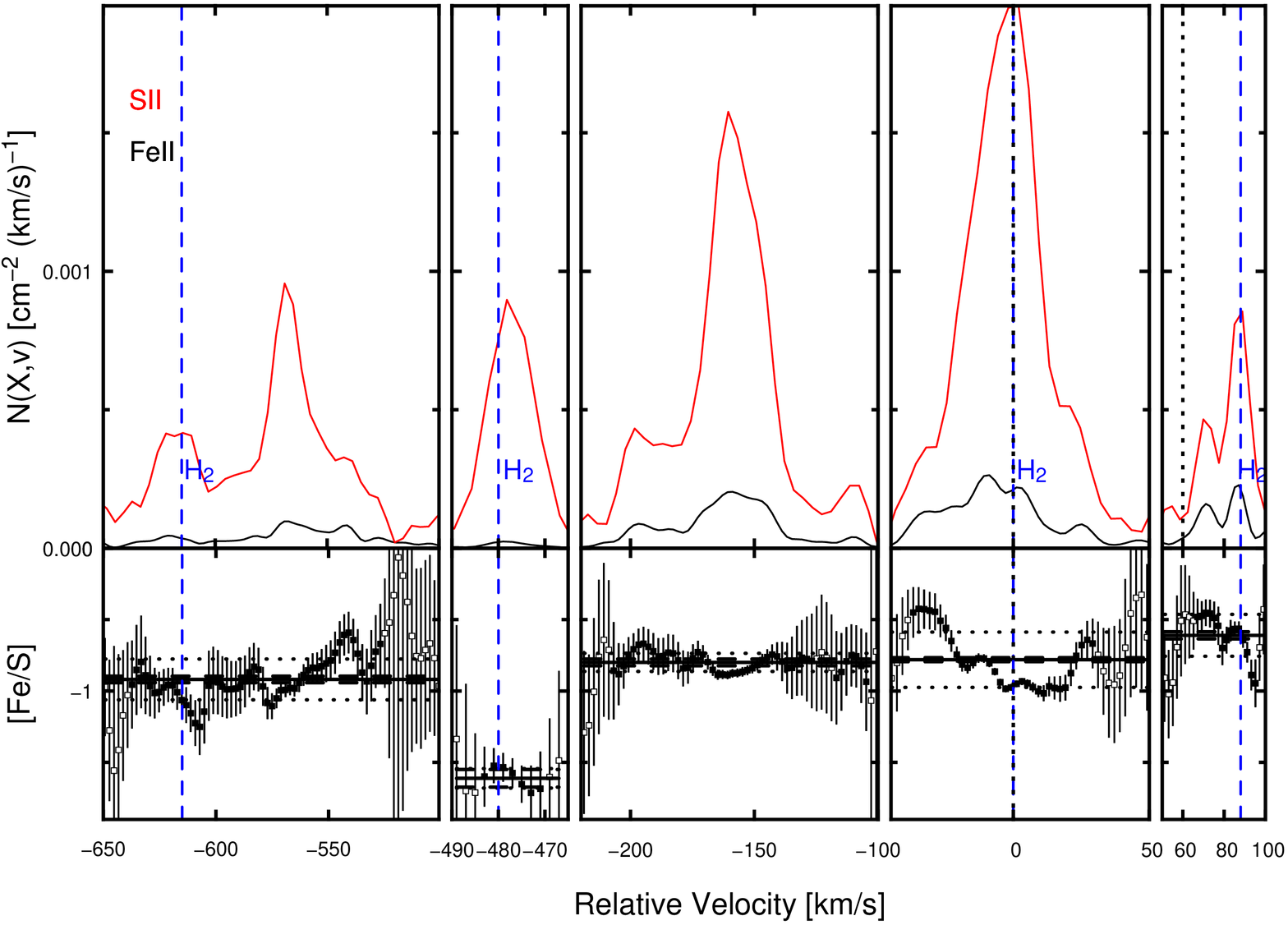}
  \caption{In this figure, top panels  show the same Fe and S profiles
    than Fig.~\ref{fig:q0013a}, divided in five regions. Lower pannels show the ratio [Fe/S] computed for each pixel. Points with errors that are too big are rejected to
    compute the mean, represented here as empty squares, while
    valid points are represented as filled squares. The mean value is plotted as a solid
    line. Dashed lines represent the $\pm\sigma$ level, and the dotted
    line the typical scatter of the points. 
H$_2$ is detected in four components at $\sim$$-$615, 
$-$480, 0, and 85~km~s$^{-1}$.}
  \label{fig:q0013b}
\end{figure*}

As mentioned in Sect.~\ref{subsec:q0013_desc}, the absorption
system found in the line of sight of Q~0013$-$004 at $z_{\rm abs}=1.973$
shows a complex multicomponent structure. Besides the fact that the
system is composed of a DLA  at  $z_{\rm abs}=1.96753$ and a sub-DLA at 
$z_{\rm abs}=1.9733$ (Petitjean et al. 2002), the low-ionization 
absorptions span about 1000~km~s$^{-1}$, and the system exhibits 
four molecular components  at $\sim$$-$615, 
$-$480, 0 and 85~km~s$^{-1}$ relative to the central redshift.
Fig.~\ref{fig:q0013a} illustrates this complexity. The 
apparent column density per velocity bin, referred to solar values, is
plotted for Fe~{\sc ii}, S~{\sc ii}, and Si~{\sc ii}. 
All have complex but remarkably similar profiles.
The bottom panels represent the ratio per velocity bin [Fe/S] for
different velocity ranges. 
The mean values (see Table~\ref{tab:res}) are smaller than in
other systems but are again quite similar 
from one sub-clump to the other.
Within a peculiar sub-clump, scatter is small, on the order of
0.2~dex. In the components where H$_2$ is detected, 
the depletion factor is of the order on [Fe/S]~=~$-$1,
therefore slightly smaller but not much less than in the overall system,
except in the component at $-$480~km~s$^{-1}$, which is the 
only region in DLAs known to date where depletion is similar to the cold gas of the Galactic disc.

\subsection{Q~0405$-$443}


\begin{figure*}
  \centering
  \includegraphics[height=12cm,width=7.5cm,angle=270]{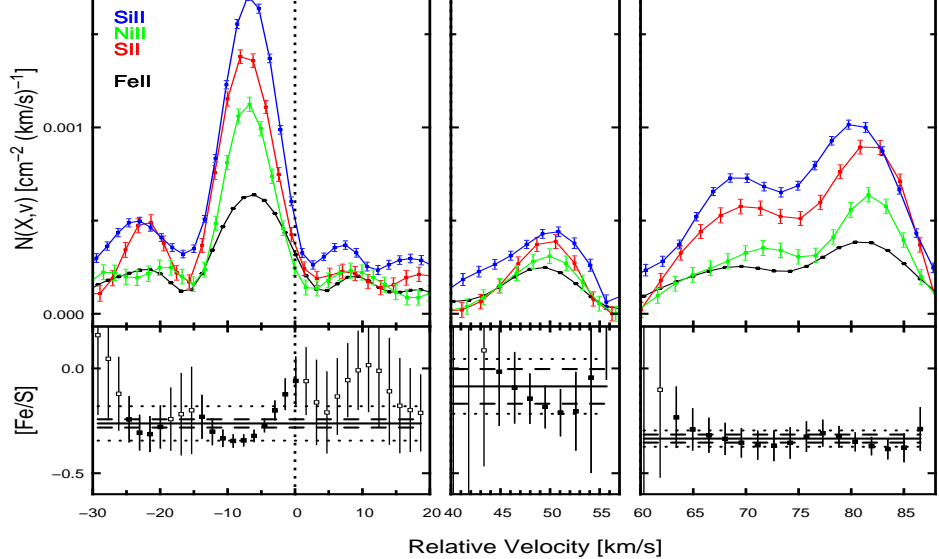}
  \caption{Top panel: optical depth profiles in 
the $z_{\rm abs}=2.549$ DLA toward
    Q~0405$-$443. The apparent column density per velocity bin -$N(X,dv)$- 
is represented on a velocity scale for Si, Ni, S, and Fe; with $v=0$~km~s$^{-1}$ centered
    at $z_{\rm abs}=2.549$. The lower panels show the ratio [Fe/S]
computed for each pixel. The mean value is plotted as a solid
    line. Dashed lines represent the $\pm\sigma$ level and the dotted
    line the typical scatter of the points.
}  
  \label{fig:q0405b_diff_rapp}
\end{figure*}

\begin{figure}[tbh]
  \centering
\includegraphics[height=10cm,width=8.0cm]{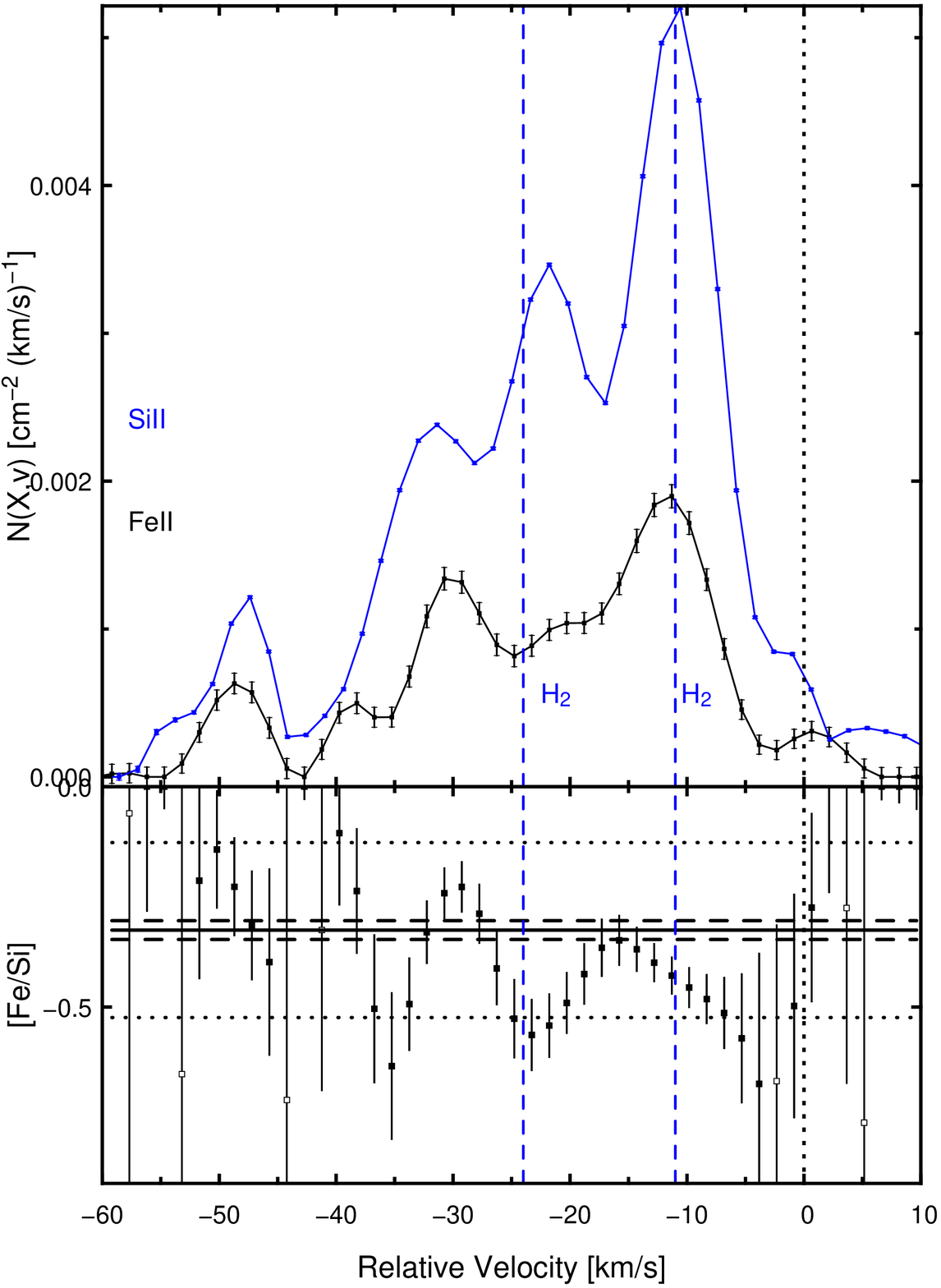}
  \caption{Top panel: optical depth profiles in 
the $z_{\rm abs}=2.595$ DLA toward
    Q~0405$-$443. The apparent column density per velocity bin -$N(X,dv)$- 
is represented on a velocity scale, with $v=0$~km~s$^{-1}$ centered
    at $z_{\rm abs}=2.595$. The lower panels show the ratio [Fe/Si]
computed for each pixel. The mean value is plotted as a solid
    line. Dashed lines represent the $\pm\sigma$ level and the dotted
    line the typical scatter of the points.
This DLA has two molecular
  components at $v\sim-24$ and $\sim-11$~km~s$^{-1}$.
}
  \label{fig:profiles_q0405a}
\end{figure}

We analyzed the systems at $z_{\rm abs}=2.549$ and
$z_{\rm abs}=2.595$ toward Q~0405$-$443. 
In the latter system,
molecular hydrogen has been detected with a column density,
log~$N$(\hdos~)~=~18.16 \citep{Ledouxsurvey}. Relatively narrow,
low-ionization transitions span about 70~km~s$^{-1}$. 
It is apparent from Fig.~\ref{fig:profiles_q0405a} that there
are some inhomogeneities inside the system and there is more
depletion (by about a factor of two) in the two components where 
H$_2$ is detected at $v$~$\sim$~$-$8 and $-$24~km~s$^{-1}$. 
Note that the depletion is larger in these components, but not 
by a large amount and definitively not as in cold gas of the
Galaxy disc.
The scatter of the pixel values is much larger than the errors 
for all element ratios (see  Table~\ref{tab:res}). 
The overall depletion pattern is again similar to that of warm gas of 
the halo, with  a slight enhancement of Cr. 

The system at $z_{\rm abs}=2.549$, presented in
Fig.~\ref{fig:q0405b_diff_rapp}, shows absorptions spread
over  $\sim120$~km~s$^{-1}$, and is well-structured in three subclumps.
From both Fig.~\ref{fig:q0405b_diff_rapp} 
and Table~\ref{tab:res}, it can be seen that the three subclumps at 
velocities [-30,20], [40, 60], and [60, 90] are quite homogeneous but have 
slightly different mean depletion factors, [Fe/S]~=~$-$0.26, $-$0.08, and
$-$0.33, respectively.  However again, the differences are smaller
than 0.2~dex.
In addition, the depletion values are close to those 
of  the $z_{\rm abs}=2.595$ system, which is located $\sim-$3000~km~s$^{-1}$ away.

\subsection{Q~0528$-$250} 

\begin{figure*}
  \centering
  \includegraphics[height=12cm,width=7.5cm,angle=270]{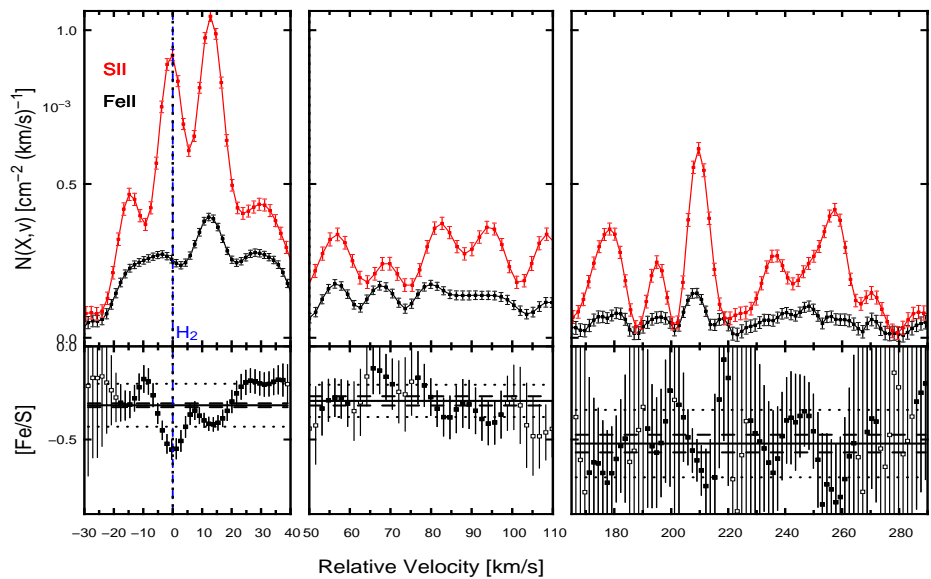}
  \caption{Top panel: optical depth profiles in 
    the $z_{\rm abs}=2.811$ DLA toward
    Q~0528$-$250. The apparent column density per velocity bin -$N(X,dv)$- 
    is represented on a velocity scale, with $v=0$~km~s$^{-1}$ centered
    at $z_{\rm abs}=2.811$. The lower panels show the ratio [Fe/S] 
    computed for each pixel. The mean value is plotted as a solid
    line. Dashed lines represent the $\pm\sigma$ level and the dotted
    line the typical scatter of the points. H$_2$ is detected in two 
    components at $\sim$~0~km~s$^{-1}$.}
  \label{fig:q0528_rappFeS}
\end{figure*}

This system is spread over more than 350~km~s$^{-1}$ and 
exhibits two molecular components located at $\sim$~0~km~s$^{-1}$. 
We observe in Fig.~\ref{fig:q0528_rappFeS}
that the internal structure is complex with many components. Note that
the profiles of Fe and S follow each other remarkably well over
most of the profile. The analysis along the profile gives
similar results for each of the subclumps,
suggesting that once again the mixing of heavy element must have been
very efficient in this system (see Table~\ref{tab:res}). 
The central part at [0, 20] has a larger depletion coefficient,
down to [Fe/S]~=~$-$0.55, when the rest of the system has
[Fe/S]~$\sim$~$-$0.3. This is the place where, again, H$_2$ is detected.
Even though the depletion factor is larger in this component, it is much
smaller than in the cold gas of the Galactic ISM.
The overall depletion pattern in the system is very close to what
is observed in the Galactic halo (see Table~\ref{tab:res}). 

\subsection{Q~1037$-$270}

\begin{figure}[tbh]
  \centering
  \includegraphics[height=10cm,width=8.0cm]{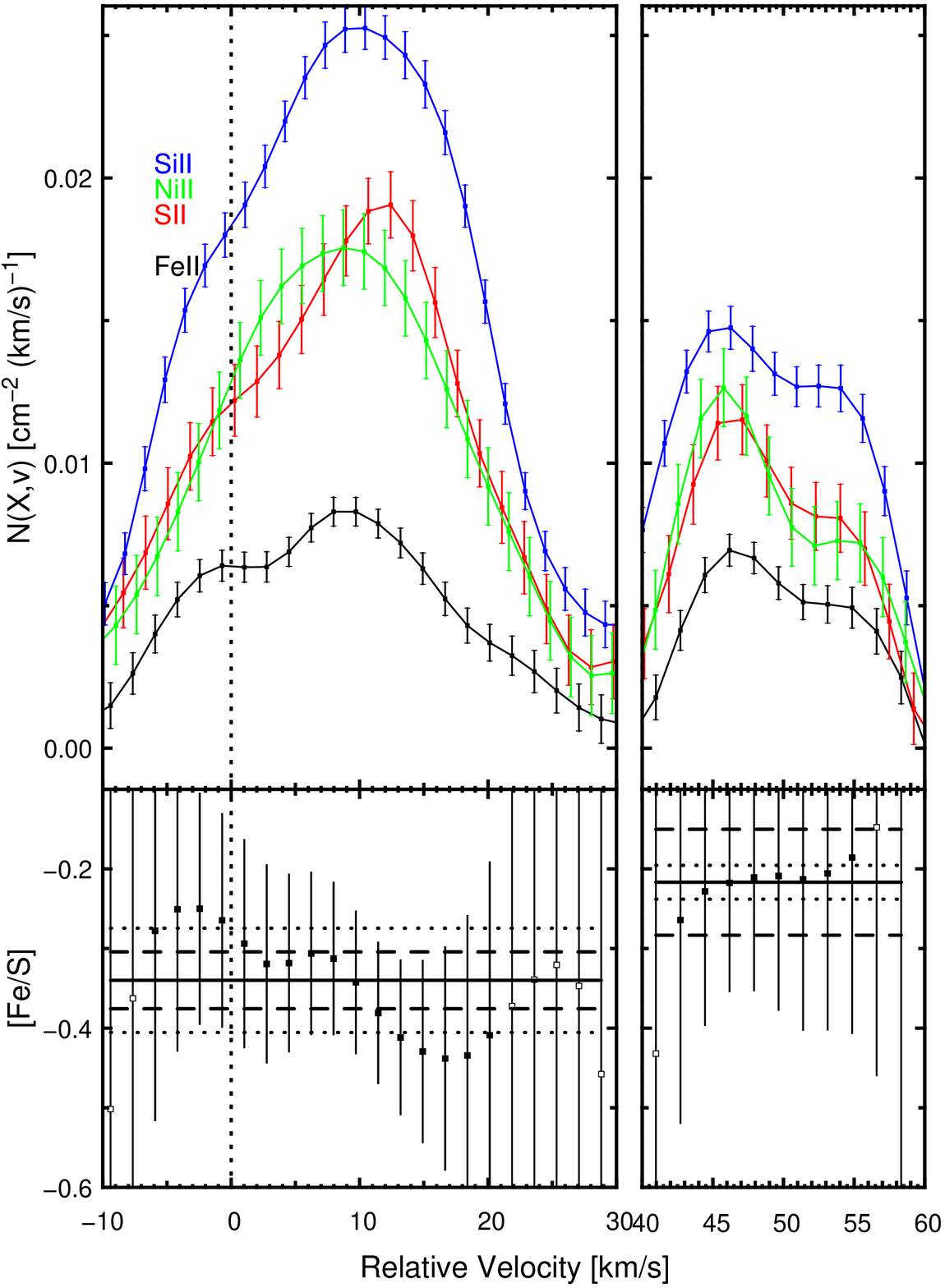}
  \caption{Top panel: optical depth profiles
in the $z_{\rm abs}=2.139$ DLA towards
    Q~1037$-$270. The apparent column density per velocity bin -$N(X,dv)$- 
is represented on a velocity scale, with $v=0$~km~s$^{-1}$ centered
    at $z_{\rm abs}=2.139$. The lower panel shows the ratio [Fe/S] 
computed for each pixel. The mean value is plotted as a solid
    line. Dashed lines represent the $\pm\sigma$ level and the dotted
    line the typical scatter of the points.} 
  \label{fig:q1037_diff_rapp}
\end{figure}

Figure~\ref{fig:q1037_diff_rapp} shows the column density per pixel
along the profile for different species, including Fe~{\sc ii}, Si~{\sc ii},
S~{\sc ii}, and Ni~{\sc ii}. All species follow the same pattern, and their 
ratios remain fairly constant across the profile.

We considered two subclumps for $-$10~$<$~$v$~$<$~30 and 40~$<$~$v$~$<$~60~km~s$^{-1}$ 
(see  Table~\ref{tab:res}). There is essentially no difference 
between the two subclumps.
Note that the scatter is comparable to or smaller than the errors
(for the first [-10,30] subclump, $\sigma_{\rm e}=0.04$,
  $\sigma_{\rm s}=0.07$ and for the second,  [40,60],
  $\sigma_{\rm e}$[Fe/S]~=~0.07 and $\sigma_{\rm s}=0.02$). 
This system is therefore fairly homogeneous within 0.1~dex,
and the abundance pattern similar to what is observed in warm gas 
of the Galactic halo.


\subsection{Q~1157+014}

\begin{figure}[tbh]
 \centering
\includegraphics[height=10cm,width=8.0cm]{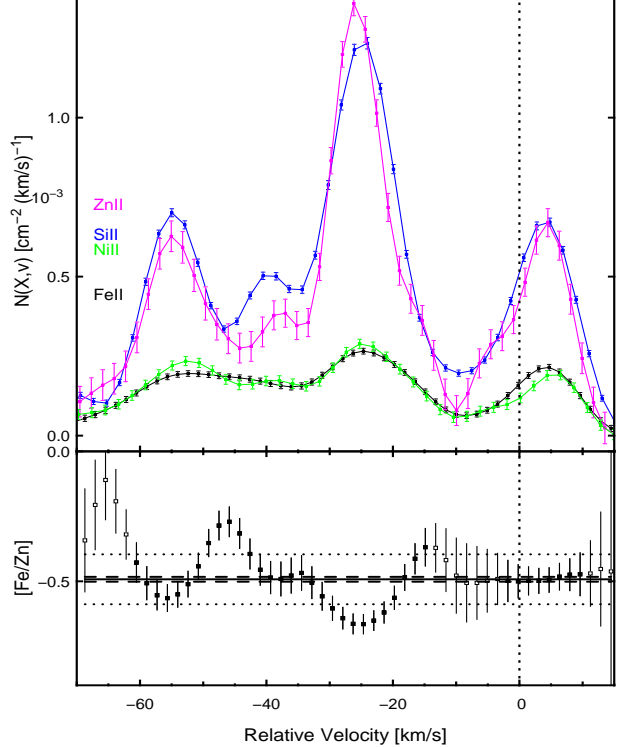}
  \caption{Top panel: optical depth profiles in 
the $z_{\rm abs}=1.944$ DLA towards
    Q~1157+014. The apparent column density per velocity bin -$N(X,dv)$- 
is represented on a velocity scale for Zn, Si, Ni, and Fe species, with $v=0$~km~s$^{-1}$ centered
    at $z_{\rm abs}=1.944$. The lower panel shows the ratios [Fe/Zn]
computed for each pixel. The mean value is plotted as a solid
    line. Dashed lines represent the $\pm\sigma$ level, and the dotted
    line the typical scatter of the points.}
 \label{fig:profiles_q1157}
\end{figure}

This system is spread over about 100~km~s$^{-1}$ and no 
associated molecules were detected. We considered only one clump.
Pixel analysis is shown in Fig.~\ref{fig:profiles_q1157}. 
Unfortunately, S~{\sc ii} lines lie at the bottom of a BAL trough, 
so we used zinc as the non-depleted species reference
instead.

The system has a low mean metallicity, [Zn/H]~=~$-$1.40
(see Table~\ref{tab:obs}), and shows a depletion pattern similar to the halo
of the galaxy for every element analyzed. 
Pixel analysis is shown in Fig.~\ref{fig:profiles_q1157}.
It can be seen that depletion is larger, [Fe/Zn]~$\sim$~$-$0.65, 
in the strongest absorption component. It should be noticed that 
21~cm absorption has been reported by \citet{Kanekar} at this
velocity ($\sim$~$-$28~km~s$^{-1}$), revealing dense gas.
The difference in depletion between the strongest subcomponent 
(at $-$28~km~s$^{-1}$) and the gas at $v$~$\sim$~$-$40~km~s$^{-1}$ is
significant (about a factor of two larger than 3$\sigma$). This
is clearly a sign of more depletion in the dense gas producing the
21~cm absorption. The overall scatter is only about 0.12~dex
(see Table~\ref{tab:res} ). This
is fairly small compared to what is observed through the ISM of our
Galaxy \citep{Welty99}. In addition, Ni and Fe follow each other very
well.

\section{Discussion}\label{sec:discussion}

\subsection{Relative abundance ratios }\label{subsec:histo}

\begin{figure*}
\centering
\includegraphics[height=5.5cm,width=7.5cm]{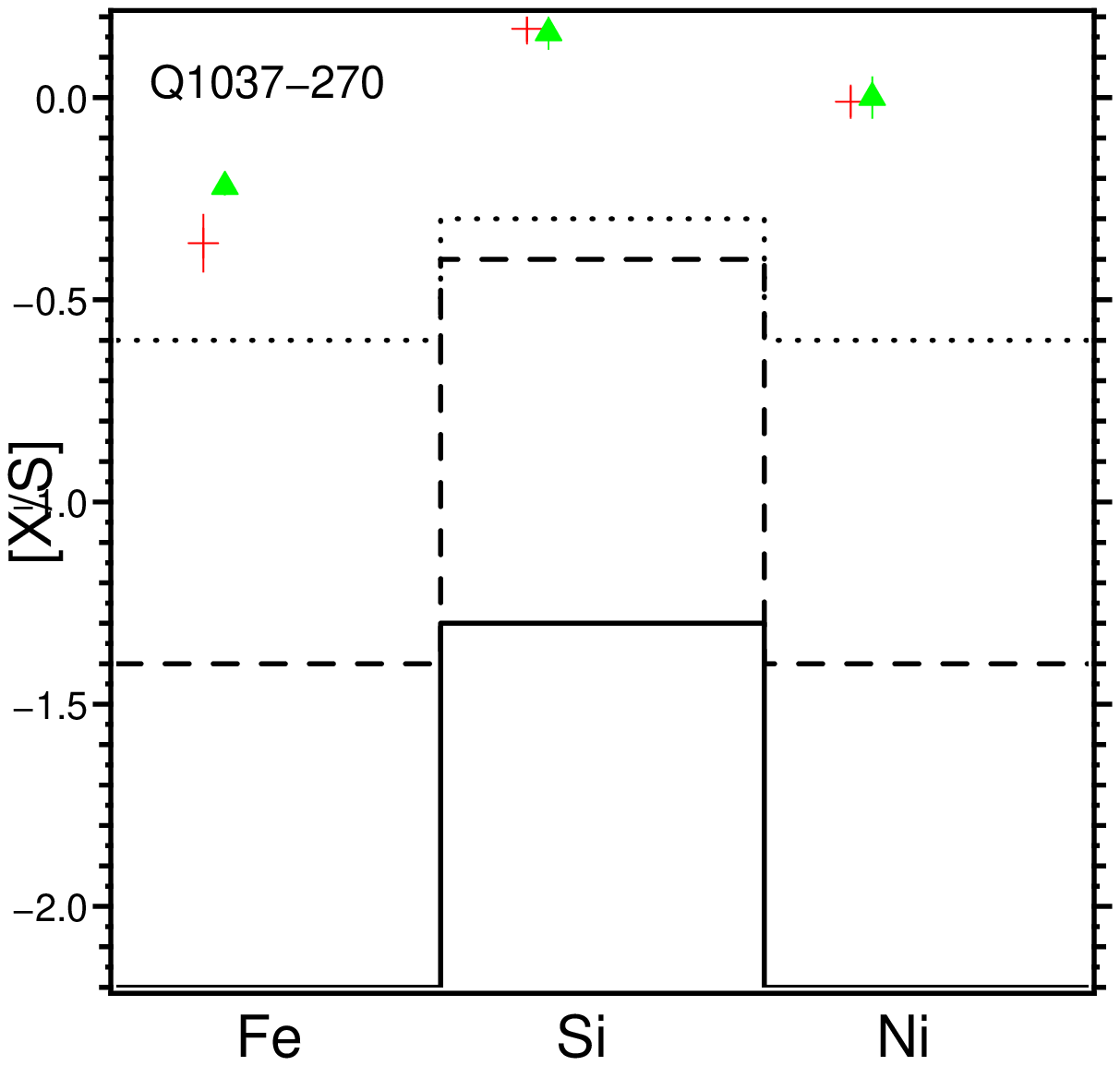}
\includegraphics[height=5.5cm,width=7.5cm]{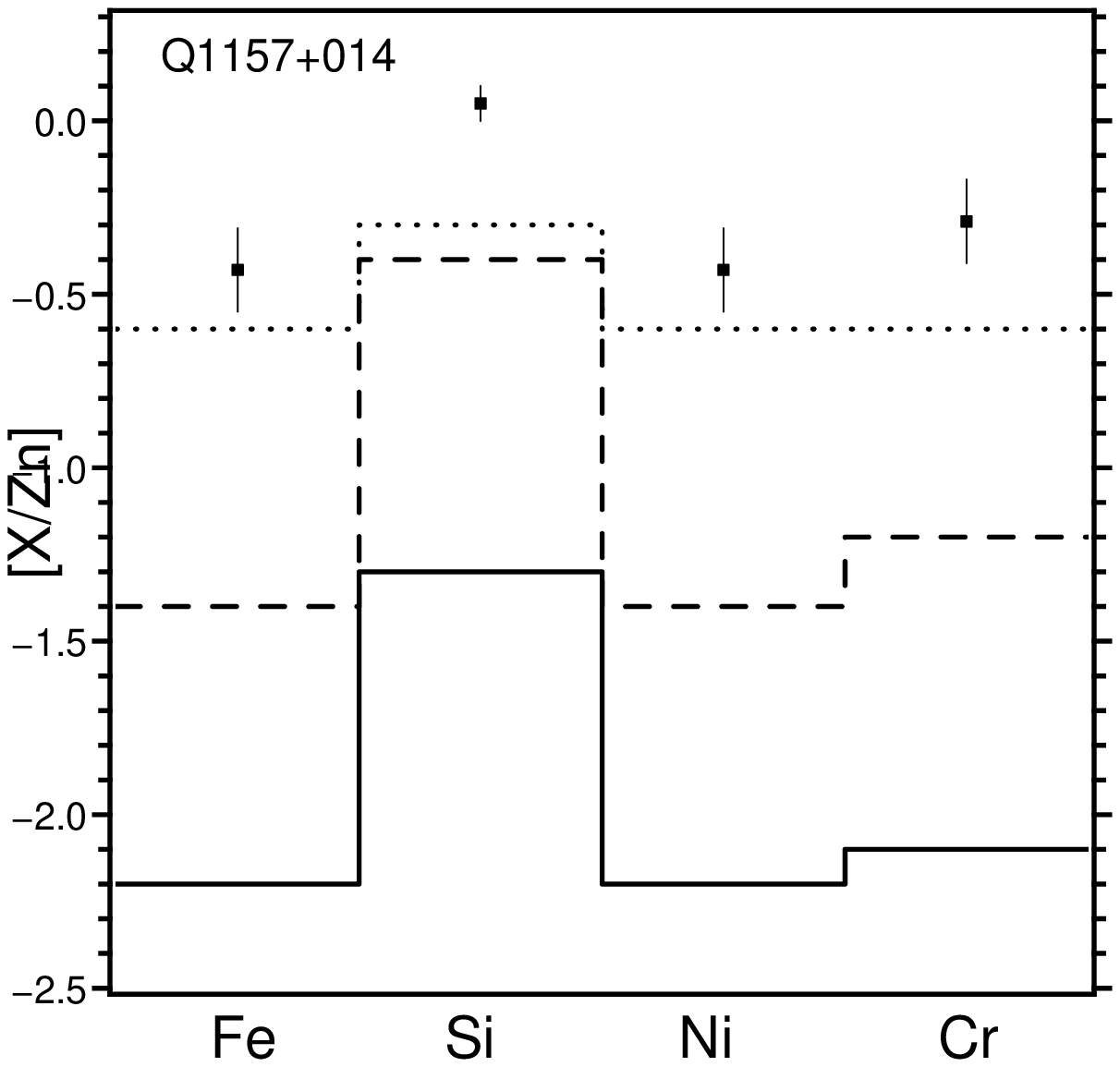}
\includegraphics[height=5.5cm,width=7.5cm]{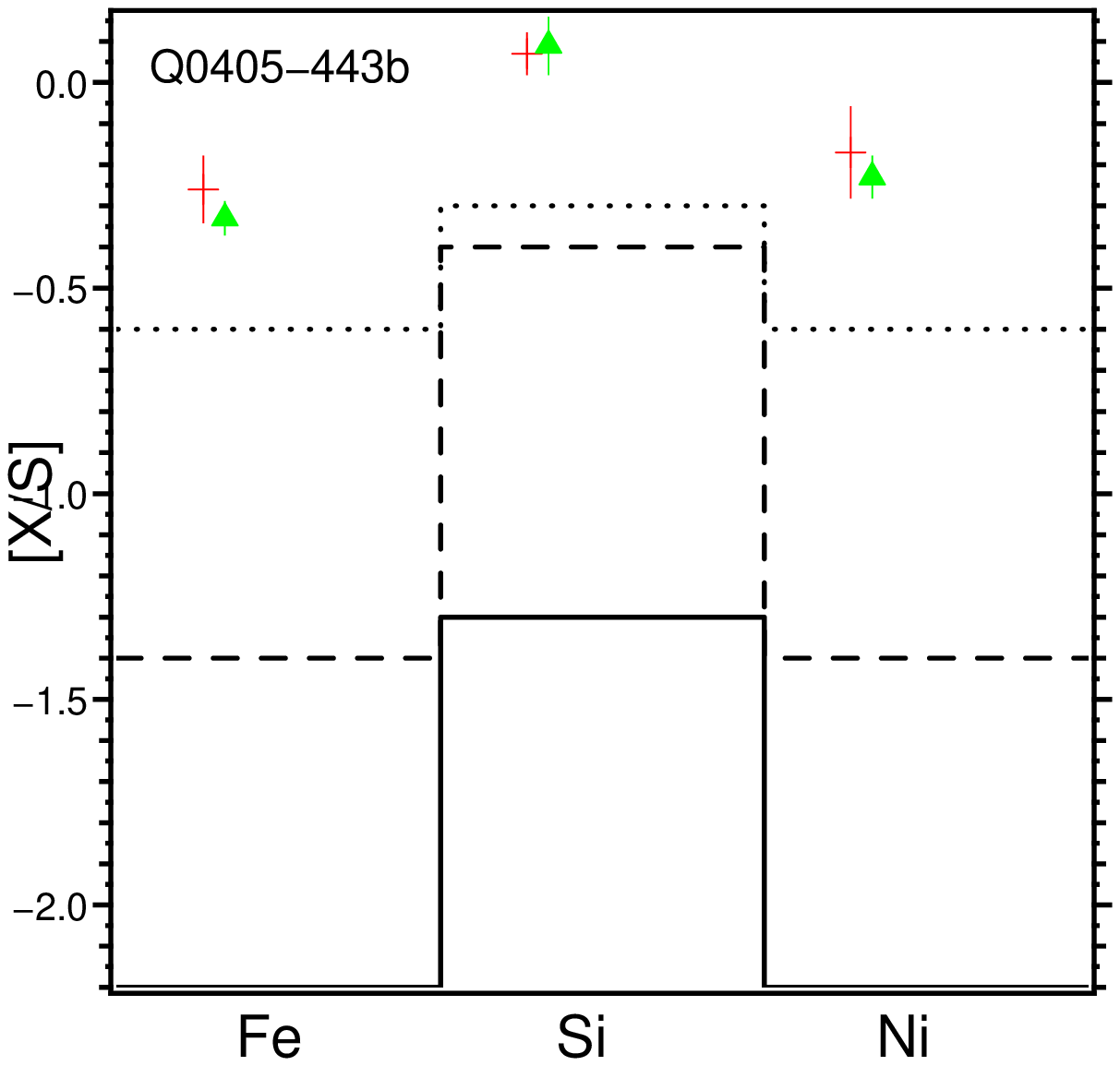}
\includegraphics[height=5.5cm,width=7.5cm]{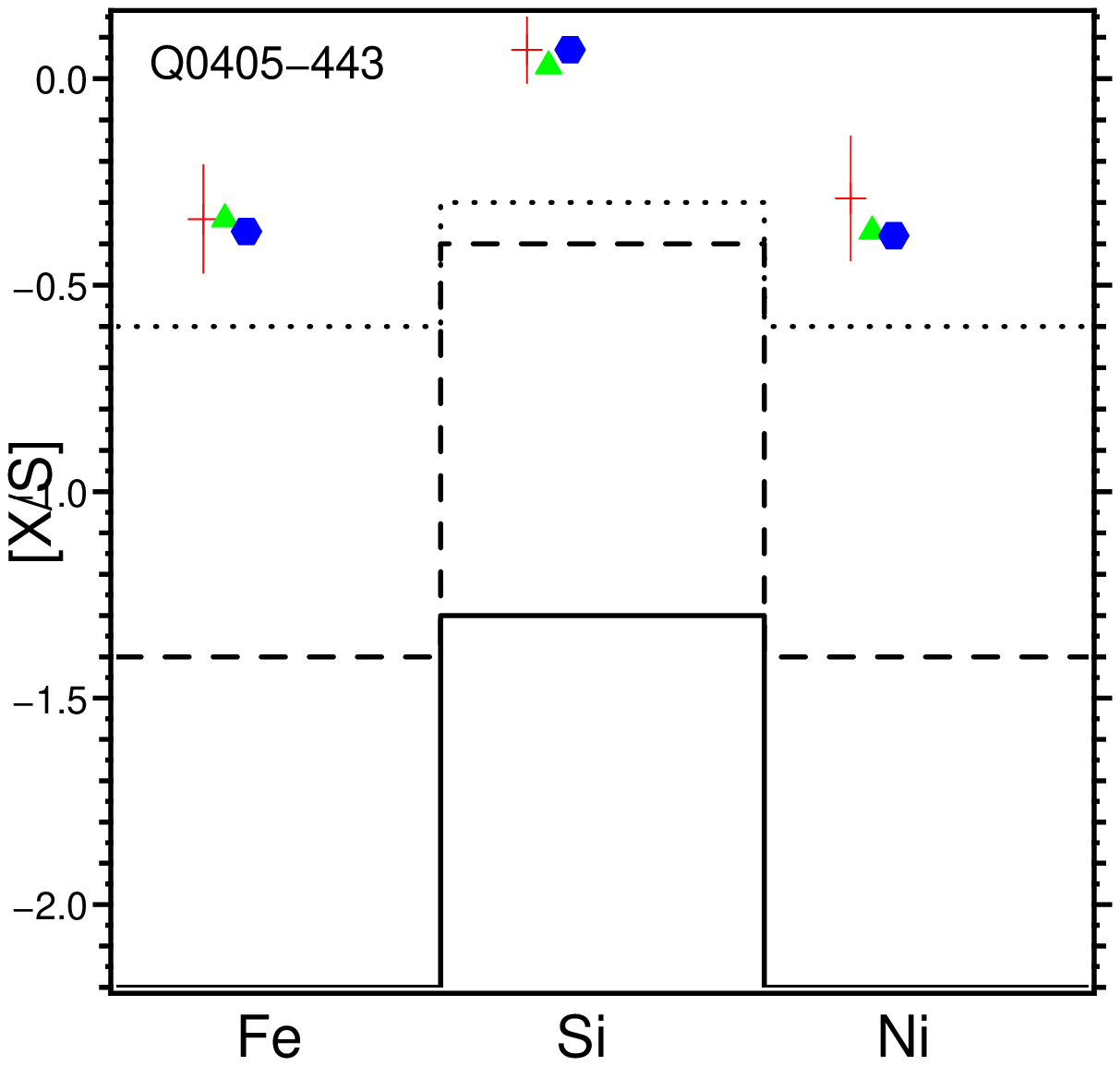}
\includegraphics[height=5.5cm,width=7.5cm]{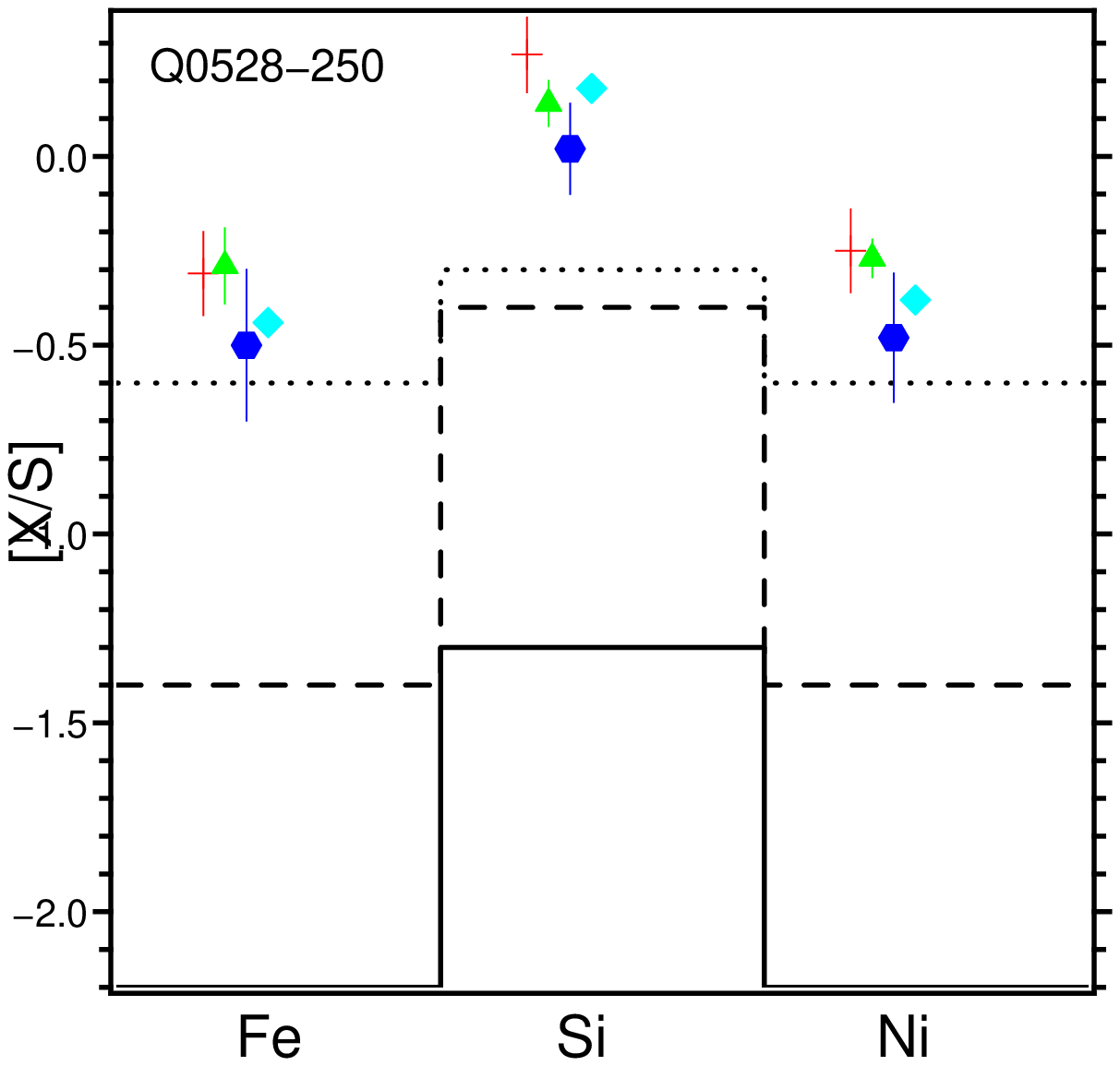}
\includegraphics[height=5.5cm,width=7.5cm]{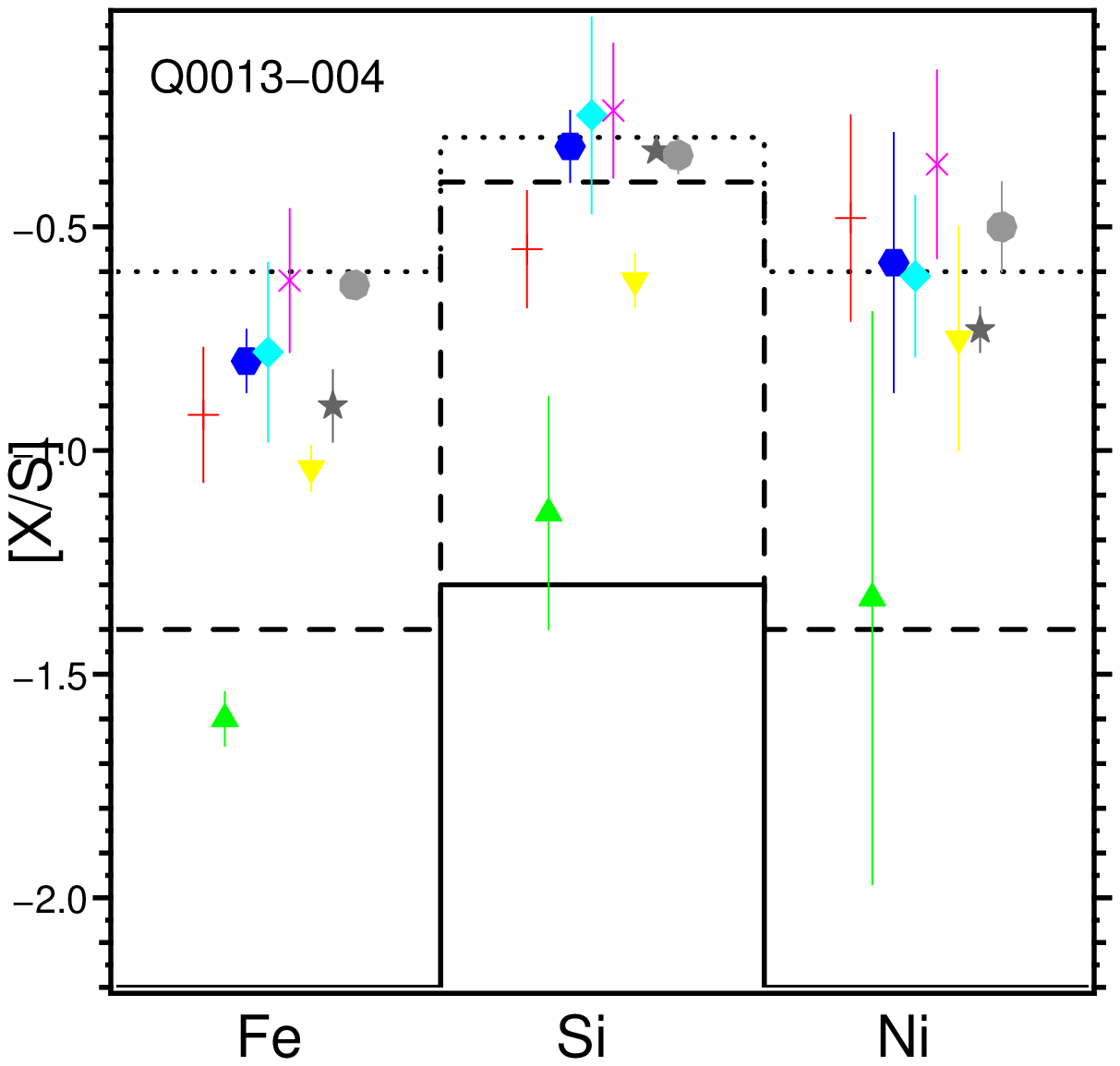}
\caption{Depletion of heavy elements relative to Sulfur (except for
  Q~1157+014, where Zinc was used instead) in the different subclumps of
the six DLA systems studied here. 
Filled symbols represent the different sub-systems considered
(see  Table~\ref{tab:res}).
Error bars
  correspond to typical scatter for each sub-system. The histograms show the
  observed values in the cold (solid line) or warm (dashed line) disc clouds and
in halo clouds (dotted line) of the Galaxy.}
\label{fig:depletion_pattern}
\end{figure*}

\begin{figure*}
\centering
\includegraphics[height=10.5cm,width=13.5cm]{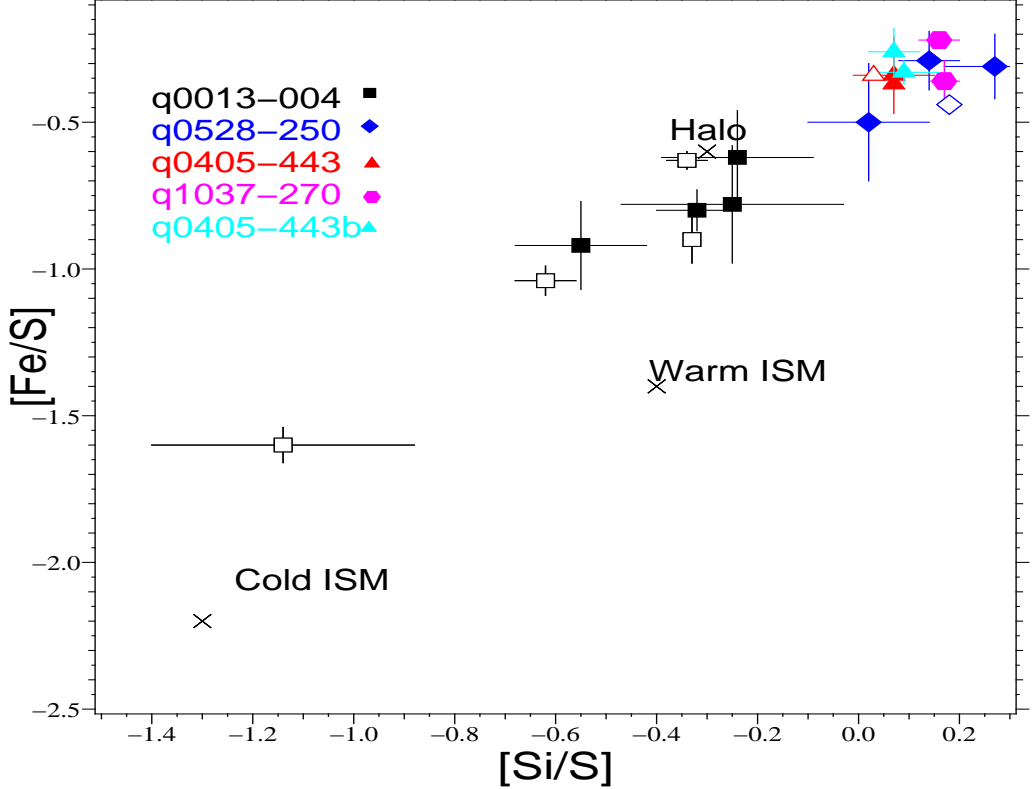}
\caption{[Fe/S] vs [Si/S] for all the subclumps analyzed in this
  paper. Different symbols represent different DLA: squares for
  Q~0013$-$004, diamonds for Q~0528$-$250, triangles for Q~0405$-$443, and
  circles for Q~1037$-$270. Open symbols are used to distinguish 
  subclumps  where H$_2$ is detected. Otherwise symbols are filled. 
We also indicate the typical [Fe/S] vs [Si/S] values 
observed in the cold,
  warm ISM, and halo of our Galaxy from \citet{Welty99}.}
\label{fig:correl}
\end{figure*}

\begin{figure*}
\centering
\includegraphics[height=5.5cm,width=8.0cm]{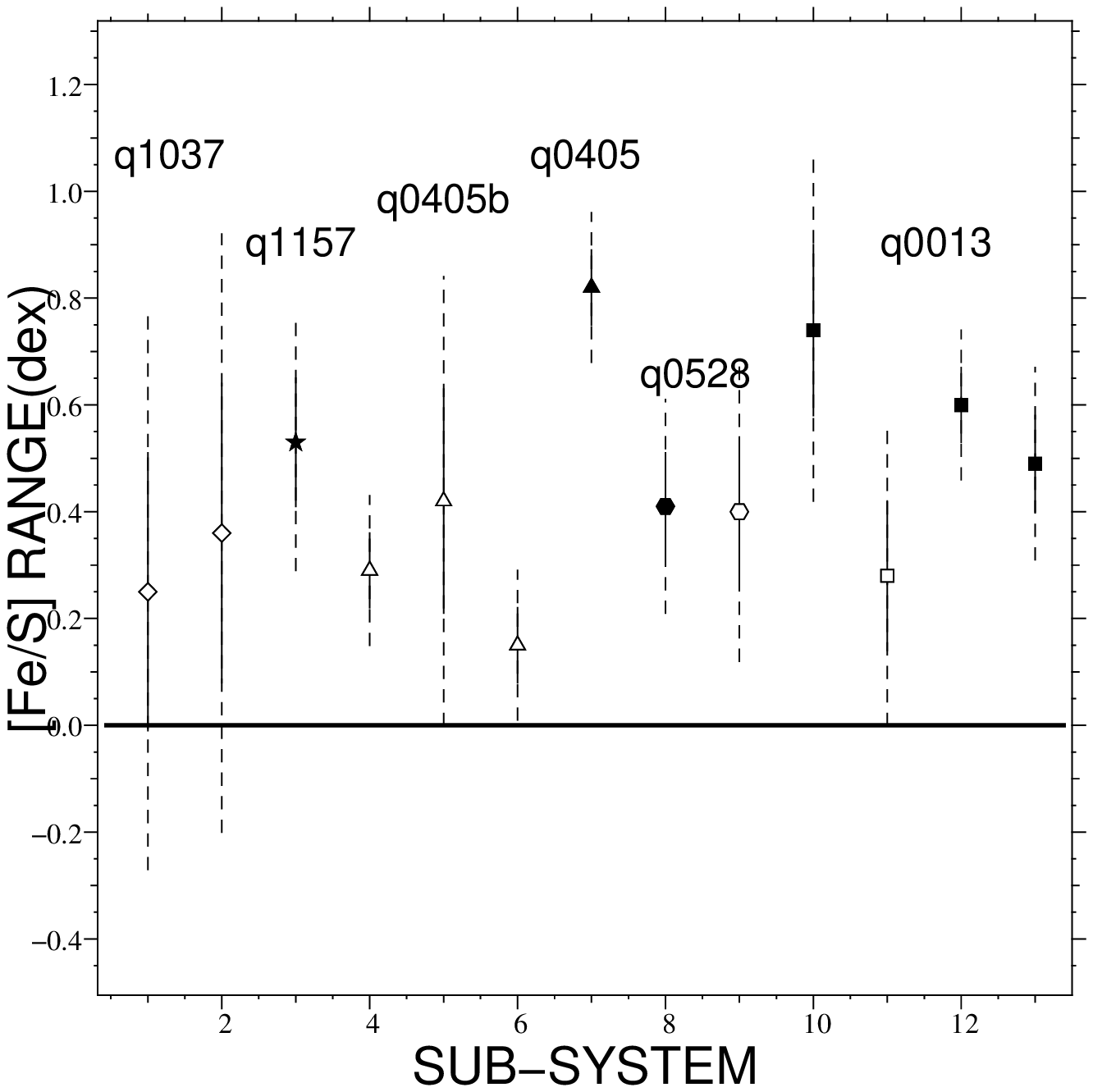}
\includegraphics[height=5.5cm,width=8.0cm]{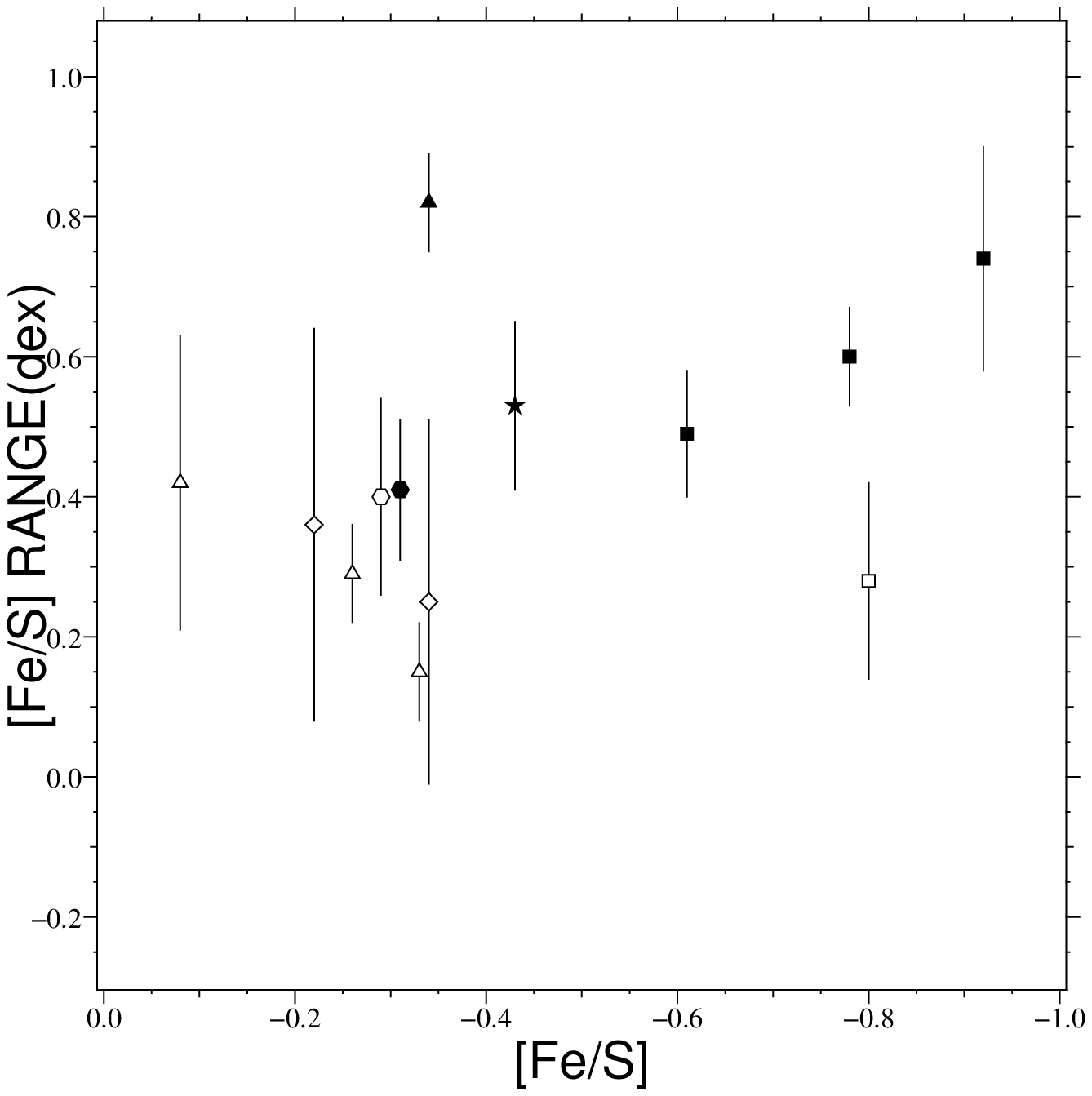}
\caption{Summary of the inhomogeneity amplitude 
observed in the subsystems. 
The y-axis represents the maximum deviation of
$[Fe/S]\pm2\sigma$ in each subclump. 
Subsystems with molecules are respresented with
filled symbols.}
\label{fig:inhomog}
\end{figure*}

Relative abundance ratios for each sub-clump considered in the six systems analyzed in this work are given in  Table~\ref{tab:res}. In the same way, results are summarized in Fig.~\ref{fig:depletion_pattern}.
It is apparent that, within each
subsystem, large departures from the mean ratio are rare: the
scatter is small if we take the observational and
fitting uncertainties into account .

Moreover, when we compare depletion values from one sub-clump to
another in the same system, differences are small. A distinct depletion pattern
is observed only for some molecular components .This is remarkably summarized by Fig.~\ref{fig:correl}, where we plot
the [Fe/S] ratio versus the [Si/S] ratio in all subclumps.
First, as already emphasized by Petitjean et al. (2002) and
Ledoux et al. (2002b), the sequence seen in this
figure is a dust-depletion sequence. Indeed, there is a correlation
between the two quantities which is expected if the
depletion is due to the presence of dust.
Secondly, the values measured in different clumps of the
same system are gathered at the same place in the figure.
The only exception is the H$_2$ component at $-$480~km~s$^{-1}$ 
in Q~0013$-$004 (see above).
Thirdly, most of the depletion pattern is similar to
that of the gas observed in the Galactic halo.
Finally, it seems that silicon is overabundant by about 0.2~dex even
relative to sulfur.
In all this, however, it must be recalled that we do not
have access to the absolute metallicity in the subclumps,
because we are not able to disentangle the H~{\sc i} 
absorptions of the different subclumps.

In the left hand panel of Fig.~\ref{fig:inhomog}, 
we plot the scatter, measured as $\sigma$,
of the ratio [Fe/S] in all the subclumps considered. In this case, we 
considered the subclumps as a whole, not isolating the molecular
component, as our aim was to see if there is a relation between the
inhomogeneity of a system and its molecular content.
The mean value of $\sigma$ over the subclumps is 0.3, which means that
inhomogeneities are less than a factor of 2. Only a few subclumps
where H$_2$ is detected have larger $\sigma$. This is expected 
because we have seen that depletion is larger over the 
specific small velocity ranges over which H$_2$ is detected.

In the right hand panel of Fig.~\ref{fig:inhomog},
we plot the different
scatter values for each subclump as a function of the total [Fe/S]
ratio. This figure confirms that (i) larger [Fe/S] ratios
are observed in subclumps where H$_2$ is detected with one
exception in a subclump of Q~0013$-$0004, and that (ii) the
scatter is larger for subclumps where H$_2$ is detected.


\subsection{The presence of \hdos~ molecules}\label{subsec:molecules}

\cite{Ledouxsurvey} have systematically  searched for
molecular hydrogen in high redshift DLAs, with a $\sim$20\% detection rate
over the whole sample.
The observed molecular fraction is often much smaller than in the ISM
of the Galactic disc (Rachford et al. 2002) and is closer to
what is observed in the magellanic clouds (Tumlinson et al. 2002).  
Here, we confirm what was already noticed by Ledoux et al. (2002b)
and \citet{ppjq0013} that,
although the presence of molecules sometimes reveals  gas with
larger depletion into dust grains than average, this is not always the case.
In most of the systems, the depletion factor is only a factor of
two larger in the components with H$_2$ compared to the overall
system. There are a few exceptions, the most noticeable being the molecular
component at $-$480~km~s$^{-1}$ toward Q~0013$-$014, in which depletion
is as large as in the cold gas of the Galactic disc.

\subsection{Consequences}\label{subsec:overall}
Variations in the relative abundance pattern within
DLAs are expected from different nucleosynthesis histories and 
from its depletion onto dust.
The magnitude of variation in the nucleosynthesis pattern may depend
on the history of star formation and the level of enrichment.
For example, stars in the Milky Way, Magellanic clouds, and local 
dwarf galaxies with metallicities varying from solar to 1/100 of solar
show a dispersion in [Si/Fe] of the order of 0.3~dex (e.g. Shetrone
et al. 2001, Venn et al. 2001). 

Peculiar nucleosynthesis histories may be reflected in the
variation of abundance ratios from one subclump to the other.
Although we observed differences in the relative metal abundances
of different sub-clumps, they are not large. This may indicate
that sub-clumps in DLAs have the same origin and history
and could be part of the same object. 
This  contrasts with the large differences in absolute metallicities 
that have been observed in LLS with similar velocity differences
(e.g. D'Odorico \& Petitjean 2001).
%
%

That depletion onto dust depends on the local physical conditions should induce a large scatter in the 
observed pixel-to-pixel relative abundance ratios. The fact that only
 small scatter was observed may reveal that the gas in DLAs is neither very 
dense nor cold but rather diffuse and warm. At least the filling factor 
of highly depleted gas is small.

All this implies 
uniform physical conditions and homogeneous and efficient mixing. One 
can speculate that this is only possible if DLAs are small objects
with dimensions on the order of one kilo-parsec. This is difficult
to ascertain as direct detection of high-redshift DLAs have not been
very successful till now (e.g. Kulkarni et al. 2000, M\o ller et al. 2004).
It is, however, very important to pursue these observations
in order to better constrain the nature and physical properties of these
objects.
%

%



\begin{acknowledgements}
RS and PPJ gratefully acknowledge support from the Indo-French
Center for the Promotion of Advanced Research (Centre Franco-Indien
pour la Promotion de la Recherche Avanc\'ee) under contract
No.3004-3. This work was supported by the European Community
Research and  Training Network: 'The Physics of the Intergalactic Medium'.
\end{acknowledgements}


\begin{table*}
\centering
\begin{small}
 \caption{This table summarizes the abundance ratios for the six systems studied. Each system was decomposed in different sub-clumps. For each range of velocities, we
 determined the mean value of the [X/S] ratio, its error $\sigma=\sqrt{\sum_{i=1}^{n}\sigma^2_{i}/n}$, as well as the
  standard deviation around the mean (in italics),as an indicator of the
  inhomogeneity through the subclump. The molecular components
 were isolated and their depletion values marked in bold face.}\label{tab:res}
\begin{tabular}{lrcccc}\hline\hline
 System    &    Sub-clump [km~s$^{-1}$] & [Fe/S]  & [Si/S] & [Ni/S] &[Cr/S] \\\hline
Q~0013$-$004& [$-$650,$-$500]&$-0.92\pm0.02\pm\textit{0.15}$&$-0.55\pm0.03\pm\textit{0.13}$&$-0.48\pm0.10\pm\textit{0.23}$& ---\\
     &$\mathbf{[-630,-605]}$&$\mathbf{-1.04\pm0.05}$&$\mathbf{-0.62\pm0.06}$&$\mathbf{-0.75\pm0.25}$&---\\
     &$\mathbf{[-490,-465]}$&$\mathbf{-1.60\pm0.06}$&$\mathbf{-1.14\pm0.26}$&---&---\\
     &[$-$220,$-$100]&$-0.80\pm0.02\pm\textit{0.07}$&$-0.31\pm0.08\pm\textit{0.08}$&---&---\\
     &[$-$45,+50]&$-0.78\pm0.01\pm\textit{0.20}$&$-0.24\pm0.06\pm\textit{0.22}$&$-0.61\pm0.06\pm\textit{0.18}$&---\\
     &$\mathbf{[-15,+15]}$&$\mathbf{-0.90\pm0.01}$&$\mathbf{-0.33\pm0.01}$&$\mathbf{-0.73\pm0.05}$&---\\
     &[+50,+100]&$-0.61\pm0.02\pm\textit{0.16}$&$-0.24\pm0.06\pm\textit{0.15}$&$-0.36\pm0.09\pm\textit{0.21}$&---\\
     &$\mathbf{[+76,+95]}$&$\mathbf{-0.64\pm0.03}$&&$\mathbf{-0.34\pm0.04}$&$\mathbf{-0.50\pm0.10}$\\\hline
Q~0405$-$443&[$-$60,+10]&$-0.27\pm0.02\pm\textit{0.13}$&$0.07\pm0.01\pm\textit{0.08}$
                                            &$-0.29\pm0.02\pm\textit{0.15}$&$-0.16\pm0.03\pm\textit{0.17}$\\
 $z_{\rm
  abs}$~=~2.595&\textbf{[$-$30,$-$14]}&$\mathbf{-0.34\pm0.01}$&$\mathbf{0.03\pm0.01}$& $\mathbf{-0.37\pm0.01}$
                                            &$\mathbf{-0.16\pm0.03}$\\ 
               &\textbf{[$-$17,$-$2]}&$\mathbf{-0.37\pm0.01}$&$\mathbf{0.07\pm0.01}$
                                            &$\mathbf{-0.38\pm0.01}$&$\mathbf{-0.23\pm0.04}$\\\hline

 Q~0405$-$443&[$-$30,+20]&$-0.26\pm0.02\pm\textit{0.08}$&$0.07\pm0.02\pm\textit{0.05}$&$-0.17\pm0.04\pm\textit{0.11}$&$0.05\pm0.06\pm\textit{0.10}$\\
$z_{\rm abs}$~=~2.549&[+40,+60]&$-0.08\pm0.08\pm\textit{0.13}$&
   $0.17\pm0.10\pm\textit{0.12}$&
   $-0.05\pm0.10\pm\textit{0.04}$&$0.00\pm0.26\pm\textit{0.18}$\\
             &[+60,+90]&$-0.33\pm0.02\pm\textit{0.04}$&
 $0.09\pm0.02\pm\textit{0.07}$& $-0.23\pm0.04\pm\textit{0.05} $
 &$-0.18\pm0.08\pm\textit{0.10}$ \\\hline

 Q~0528$-$250&[$-$30,+40]& $-0.31\pm0.01\pm\textit{0.11}$&
 $0.27\pm0.01\pm\textit{0.10}$&$-0.25\pm0.01\pm\textit{0.11}$&---\\
  $z_{\rm abs}$~=~2.811&\textbf{[$-$8,+8]}&$\mathbf{-0.44\pm0.01}$&$\mathbf{0.18\pm0.01}$&$\mathbf{-0.38\pm0.02}$&---\\
            &[+50,+110]&  $-0.29\pm0.02\pm\textit{0.10}$& $0.14\pm0.02\pm\textit{0.06}$& $-0.27\pm0.04\pm\textit{0.06}$&---\\
            & [+165,+290]&$-0.50\pm0.05\pm\textit{0.20}$&
 $-0.02\pm0.02\pm\textit{0.12}$&$-0.49\pm0.06\pm\textit{0.17}$&---\\\hline

Q~1037$-$270&[$-$10,+30]&$-0.34\pm0.04\pm\textit{0.07}$&$0.17\pm0.03\pm\textit{0.03}$&$-0.01\pm0.04\pm\textit{0.04}$&---\\

$z_{\rm
  abs}=2.139$&[+40,+60]&$-0.22\pm0.07\pm\textit{0.02}$&$0.16\pm0.05\pm\textit{0.04}$&$0.00\pm0.07\pm\textit{0.05} $&---\\\hline\hline
&& [Fe/Zn] & [Si/Zn] & [Ni/Zn] &[Cr/Zn] \\\hline

 Q~1157+014&[$-$70,+15]& $-0.43\pm0.02\pm\textit{0.12}$&
 $0.05\pm0.02\pm\textit{0.05}$&$-0.43\pm0.03\pm\textit{0.12}$&
 $-0.29\pm0.02\pm\textit{0.12}$\\
 $z_{\rm abs}=1.944$&&&&\\
\hline\hline
\end{tabular}
\end{small}
\end{table*}

\end{document}